\documentclass[11pt,a4paper]{scrartcl}
\usepackage{amsmath,amssymb}
\usepackage{braket}
\usepackage{graphicx}
\usepackage{hyperref}
\usepackage{multirow}
\usepackage{color}
\usepackage{siunitx}
\usepackage{subcaption}
\usepackage{tikz}
\usepackage{todonotes}

\graphicspath{{plots/}}

\newcommand{\MSbar}{\overline{\mathrm{MS}}}
\newcommand{\app}[1]{Appendix~\ref{#1}}
\newcommand{\fig}[1]{Figure~\ref{#1}}
\newcommand{\tab}[1]{Table~\ref{#1}}
\newcommand{\eq}[1]{Eq.~\ref{#1}}

\newcommand{\mkaf}{\ensuremath{M_K a_0 =
	-0.385(16)_{\textrm{stat}} (^{+0}_{-12})_{m_s}(^{+0}_{-5})_{Z_P}(4)_{r_f}}}
\newcommand{\af}{\ensuremath{a_0 = -0.154(6)_{\textrm{stat}}(^{+0}_{-5})_{m_s}
(^{+0}_{-2})_{Z_P}(2)_{r_f}\,\si{\femto\metre}}}
\begin{document}

\title{Hadron-Hadron Interactions from $N_f=2+1+1$ Lattice QCD:\\
  isospin-1 $KK$ scattering length}

\author{C.~Helmes, C.~Jost, B.~Knippschild, B.~Kostrzewa, L.~Liu, C.~Urbach,\\
  M.~Werner\\
  \small{Helmholtz Institut f{\"u}r Strahlen- und Kernphysik, University of Bonn, Bonn, Germany}\vspace*{0.3cm}\\
} 

\maketitle

\begin{abstract}
  \begin{center}
    \includegraphics[draft=false,width=.2\linewidth]{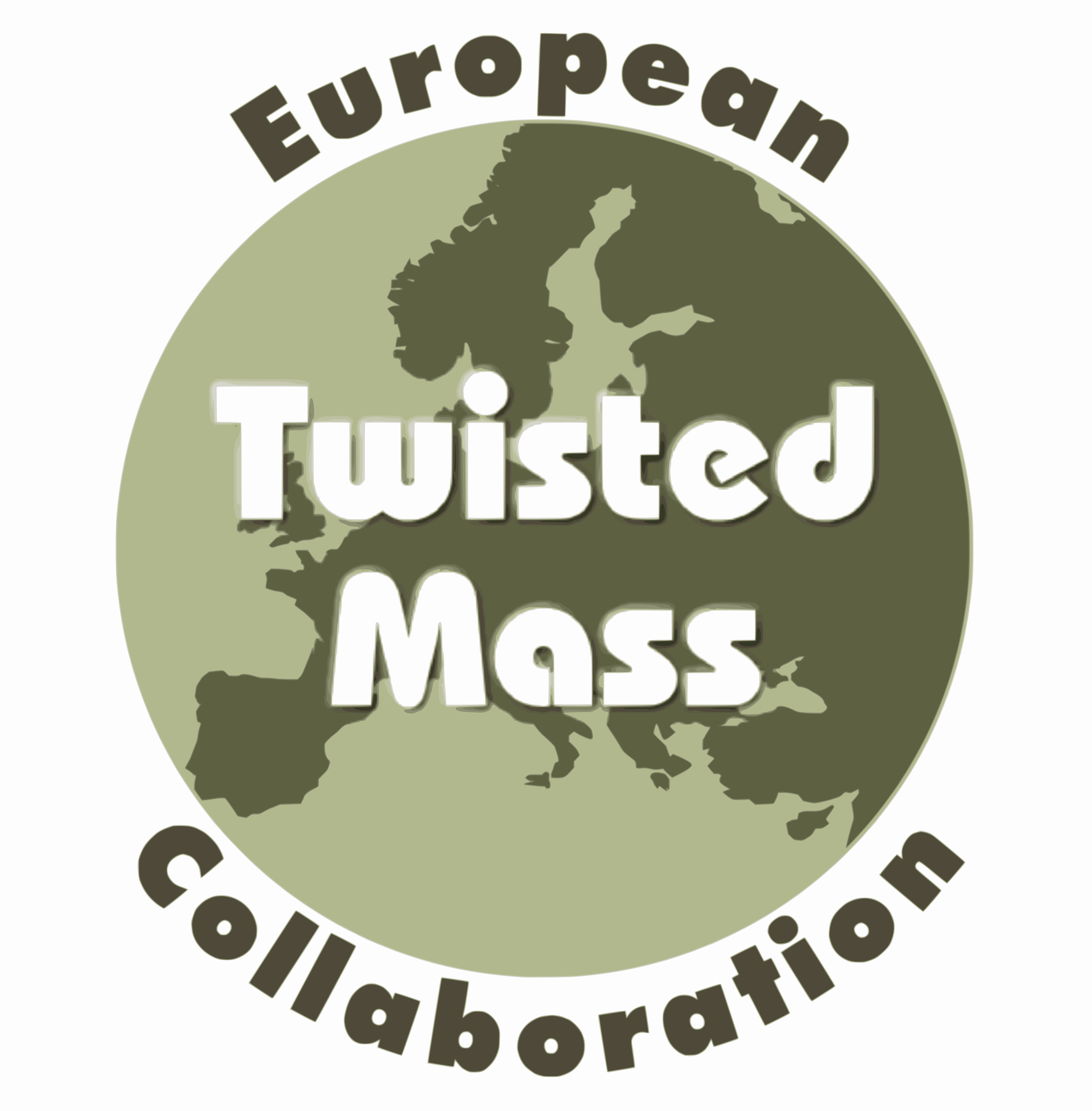}
  \end{center}
  \vspace*{0.3cm}
  We present results for the interaction of two kaons at maximal
  isospin. The calculation is based on $N_f=2+1+1$ flavour gauge
  configurations generated by the European Twisted Mass
  Collaboration with pion masses ranging from about
  \SIrange{230}{450}{\mega\electronvolt} at three values of 
  the lattice spacing. The elastic scattering length
  $a_0^{I=1}$ is calculated at several values of the bare
  strange and light quark masses. We find $\mkaf$
  as the result of a combined extrapolation to the continuum and to
  the physical point, where the first error is statistical, and the
  three following are systematical. This translates to $\af$.
\end{abstract}

\clearpage

\section{Introduction}

Shortly after the Big Bang the universe is believed to have been in a quark
gluon plasma state of matter. Apart from the inside of neutron stars the only
places where this state of matter appears and can be studied are detectors
investigating heavy ion or proton-proton collisions like the STAR detector at 
the Relativistic Heavy Ion Collider (RHIC) at
BNL~\cite{Adamczyk:2013wqm} or the ALICE experiment at the LHC at
CERN~\cite{Adam:2015vja}. The collisions 
taking place at such sites yield in their final states numerous light hadrons
like pions and kaons. Due to the mass difference between kaons and pions the
produced kaons carry much lower momenta than the pions, therefore
being much more likely to interact elastically. The interaction of two
kaons is determined by Quantum Chromodynamics (QCD), which is
non-perturbative at low energies. The understanding and interpretation
of the results of the aforementioned experiments make a
non-perturbative investigation of kaon-kaon interactions highly
desirable. While this can be formulated in chiral
perturbation theory (ChPT), it is theoretically interesting to check if
the effective approach is able to properly describe kaon-kaon scattering.
Lattice QCD provides a non-perturbative ab initio method to perform such a
study.

Hadron-hadron scattering has become more and more accessible to
lattice QCD simulations over the last years. This is on the one hand
due to L{\"u}schers finite volume formalism, and on the other hand due
to lattice QCD ensembles becoming ever more realistic.
For kaon-kaon scattering in the isospin-1 channel only a few lattice QCD
calculations have been performed~\cite{Beane:2007uh,Sasaki:2013vxa} where the
result of the former calculation has been used in
Ref.~\cite{Adam:2015vja} for the ALICE results.
In the maximal isospin channel kaon-kaon scattering resembles the well studied
pion-pion case
\cite{Yamazaki:2004qb,Beane:2005rj,Beane:2007xs,Feng:2009ij,Yagi:2011jn,Fu:2013ffa,Sasaki:2013vxa}:
there are no fermionic disconnected diagrams and only one light
quark is replaced by a strange quark. Since we already investigated
pion-pion scattering in the isospin-2 channel~\cite{Helmes:2015gla} a
lot of our analysis tools can be carried over to the present
investigation. 

In this paper we present the first study of $K^+K^+$
scattering from lattice QCD based on $N_f=2+1+1$ ensembles of the
European Twisted Mass Collaboration
(ETMC)~\cite{Baron:2010bv,Baron:2010th} covering 
three values of the lattice spacing. These ensembles, which employ up 
to five values of the light quark mass per lattice spacing value allow
us to perform reliable chiral and continuum extrapolations of our
results. 

For the strange quark we employ a mixed action
approach with so-called Osterwalder-Seiler valence quarks on the
Wilson twisted mass sea~\cite{Frezzotti:2003xj}. This allows us to tune the
valence strange quark mass value to its physical value without spoiling
the automatic $\mathcal{O}(a)$-improvement guaranteed by Wilson twisted mass
lattice QCD at maximal twist~\cite{Frezzotti:2003ni}.
However, while unitarity
breaking effects vanish in the continuum limit, this ansatz also
introduces partial-quenching effects, which we cannot control in the
present calculation. However, in previous calculations with this
setup, no sizable effects were found, see
e.g.~\cite{Farchioni:2010tb,Carrasco:2014cwa}. The mixed-action 
approach for the strange quark also allows us to avoid the parity-flavour
mixing present in the $1+1$ (strange-charm) sea sector of Wilson twisted mass
lattice QCD at maximal twist with $N_f=2+1+1$ flavours. 

Our final result differs by about $2\sigma$ from the
determinations by NPLQCD~\cite{Beane:2007uh} and about $4\sigma$ from
the determination of PACS-CS~\cite{Sasaki:2013vxa}.
This deviation can likely be attributed to
lattice artefacts: NPLQCD works mainly at a single lattice spacing
with the exception of one ensemble at a second lattice spacing
value. PACS-CS works at a single lattice spacing only. However, we can
also not exclude residual unitarity breaking effects in our
calculation. Interestingly, our result is actually equal to the
leading order ChPT prediction for $M_K a_0$.

\section{Lattice action}
\label{sec:actions}

\begin{table}[t!]
 \centering
 \begin{tabular*}{.9\textwidth}{@{\extracolsep{\fill}}lcccccc}
  \hline\hline
  ensemble & $\beta$ & $a\mu_\ell$ & $a\mu_\sigma$ & $a\mu_\delta$ &
  $(L/a)^3\times T/a$ & $N_\mathrm{conf}$  \\ 
  \hline\hline
  $A30.32$   & $1.90$ & $0.0030$ & $0.150$  & $0.190$  & $32^3\times64$ & $263$  \\
  $A40.20$   & $1.90$ & $0.0040$ & $0.150$  & $0.190$  & $20^3\times48$ & $268$  \\
  $A40.24$   & $1.90$ & $0.0040$ & $0.150$  & $0.190$  & $24^3\times48$ & $386$  \\
  $A40.32$   & $1.90$ & $0.0040$ & $0.150$  & $0.190$  & $32^3\times64$ & $244$  \\
  $A60.24$   & $1.90$ & $0.0060$ & $0.150$  & $0.190$  & $24^3\times48$ & $314$  \\
  $A80.24$   & $1.90$ & $0.0080$ & $0.150$  & $0.190$  & $24^3\times48$ & $305$  \\
  $A100.24$  & $1.90$ & $0.0100$ & $0.150$  & $0.190$  & $24^3\times48$ & $308$  \\
  \hline                                                                   
  $B35.32$   & $1.95$ & $0.0035$ & $0.135$  & $0.170$  & $32^3\times64$ & $235$ \\
  $B55.32$   & $1.95$ & $0.0055$ & $0.135$  & $0.170$  & $32^3\times64$ & $293$ \\
  $B85.24$   & $1.95$ & $0.0085$ & $0.135$  & $0.170$  & $32^3\times64$ & $290$ \\
  \hline                                                                   
  $D30.48$ & $2.10$ & $0.0030$ & $0.120$ & $0.1385$ & $48^3\times96$ & $   369$ \\
  $D45.32sc$ & $2.10$ & $0.0045$ & $0.0937$ & $0.1077$ & $32^3\times64$ & $283$ \\
  \hline\hline

 \end{tabular*}
 \caption{The gauge ensembles used in this study. For the labelling of
   the ensembles we adopted the notation in
   Ref.~\cite{Baron:2010bv}. In addition to the relevant input
   parameters we give the lattice volume and  the number of evaluated
   configurations, $N_\mathrm{conf}$.}
 \label{tab:setup}
\end{table}

We use gauge configurations generated by the ETM
collaboration with $N_f=2+1+1$ dynamical quark
flavours~\cite{Baron:2010bv}. The Iwasaki gauge action~\cite{Iwasaki:1983ck} is used
in combination with the Wilson twisted mass fermion discretisation.
There are
three values of the lattice spacing available, with
$\beta=1.90$, $\beta=1.95$ and $\beta=2.10$ corresponding to
$a\sim0.089\ \mathrm{fm}$, $a\sim0.082\ \mathrm{fm}$ and
$a=0.062\ \mathrm{fm}$, respectively. The ensembles we used are
compiled in table~\ref{tab:setup}. The lattice scale for the ensembles
has been determined in Ref.~\cite{Carrasco:2014cwa} using
$f_\pi$. Also in Ref.~\cite{Carrasco:2014cwa} the pseudoscalar renormalisation
constant $Z_P$, the inverse of which is the quark mass renormalisation
constant in the twisted-mass approach, has been determined for each lattice spacing
and then converted to the $\overline{\textrm{MS}}$ scheme at a scale of $2\ \mathrm{GeV}$.

The computation of $Z_P$ employs the RI-MOM renormalisation scheme and further makes use of two
different methods which are labelled \textbf{M1} and \textbf{M2} by the authors. The two methods,
\textbf{M1} and \textbf{M2}, give results which differ by lattice artefacts.
As an intermediate length scale we use the Sommer parameter $r_0/a$
determined in Ref.~\cite{Baron:2010bv} for each value of the light
quark mass $m_l$ and extrapolated to the chiral limit in
Ref.~\cite{Carrasco:2014cwa}, assuming either a linear or quadratic
dependence on the light quark mass. The value of $r_0$ in fm was
determined in Ref.~\cite{Carrasco:2014cwa} using chiral perturbation
theory (ChPT) employing methods \textbf{M1} and \textbf{M2} for $Z_P$,
reading 

\begin{equation}
  \label{eq:r0}
  \begin{split}
    r_0\ &=\ 0.470(12)\ \mathrm{fm}\qquad\textbf{(M1)}\,,\\
    r_0\ &=\ 0.471(11)\ \mathrm{fm}\qquad\textbf{(M2)}\,.\\
  \end{split}
\end{equation}
We keep the two values separate here, because we will use them to
estimate systematic uncertainties. 
The values for $Z_P$, the lattice spacing $a$ and $r_0/a$ are
summarised in table~\ref{tab:r0values} for the three
$\beta$-values. For details we refer to Ref.~\cite{Carrasco:2014cwa}. 
Note that $\mu_\sigma$ and $\mu_\delta$ are kept fixed for all
$\mu_\ell$ values at $\beta=1.90$ and $\beta=1.95$. Between the two
ensembles D30.48 and D45.32sc they differ slightly.

\begin{table}
  \centering
  \begin{tabular*}{.95\linewidth}{@{\extracolsep{\fill}}lrrrr}
    \hline\hline
    $\beta$ & $Z_\mathrm{P}$\ (\textbf{M1}) & $Z_\mathrm{P}$\ (\textbf{M2}) &
    $a\ [\mathrm{fm}]$ & $r_0/a$ \\
    \hline\hline
    $1.90$  & $0.529(07)$ & $0.574(04)$ & $0.0885(36)$ & $5.31(8)$ \\
    $1.95$  & $0.509(04)$ & $0.546(02)$ & $0.0815(30)$ & $5.77(6)$ \\
    $2.10$  & $0.516(02)$ & $0.545(02)$ & $0.0619(18)$ & $7.60(8)$ \\
    \hline\hline
  \end{tabular*}
  \caption{compilation of values for the Sommer parameter $r_0/a$, the
    lattice spacing $a$ and
    $Z_\mathrm{P}$ at $2\ \mathrm{GeV}$ in the $\MSbar$ scheme
    determined with methods \textbf{M1,M2} 
    the three values of the lattice spacing. See 
    Ref.~\protect{\cite{Carrasco:2014cwa}} for details.}
  \label{tab:r0values}
\end{table}

In order to set the strange quark mass, we use $M_K$ in physical units
as input. We use $M^{\textrm{phys}}_K=\SI{494.2(3)}{\mega\electronvolt}$ corrected for
electromagnetic and isospin breaking effects~\cite{Aoki:2016frl}.

As further inputs we use the average up/down quark mass, $m_l^{\textrm{phys}} =\SI{
3.70(17)}{\mega\electronvolt}$, from Ref.~\cite{Carrasco:2014cwa}
as well as the neutral pion mass, $M^{\textrm{phys}}_{\pi^0}=\SI{134.98}{\mega\electronvolt}$~\cite{Olive:2016xmw}.

In more detail, for the sea quarks we use the Wilson twisted
mass action with $N_f=2+1+1$ dynamical quark flavours. The Dirac
operator for the light quark doublet reads~\cite{Frezzotti:2000nk} 
\begin{equation}
  D_\ell = D_\mathrm{W} + m_0 + i \mu_\ell \gamma_5\tau^3\, ,
  \label{eq:Dlight}
\end{equation}
where $D_\mathrm{W}$  denotes the standard Wilson Dirac operator and $\mu_\ell$
the bare light twisted mass parameter. $\tau^3$ and in general
$\tau^i, i=1,2,3$ represent the Pauli matrices acting in flavour
space. $D_\ell$ acts on a spinor $\chi_\ell = (u,d)^T$ and, hence, the $u$
($d$) quark has twisted mass $+\mu_\ell$ ($-\mu_\ell$).

For the heavy doublet of $c$ and
$s$ quarks~\cite{Frezzotti:2003xj} the Dirac operator is given by
\begin{equation}
  D_\mathrm{h} = D_\mathrm{W} + m_0 + i \mu_\sigma \gamma_5\tau^1 + \mu_\delta \tau^3\,.
  \label{eq:Dsc}
\end{equation}
The bare Wilson quark mass $m_0$ has been tuned to its
critical value
$m_\mathrm{crit}$~\cite{Chiarappa:2006ae,Baron:2010bv}. This
guarantees automatic order $\mathcal{O}\left(a\right)$-improvement
\cite{Frezzotti:2003ni}, which is one of the main advantages of the
Wilson twisted mass formulation of lattice QCD. For a discussion on
how to tune to $m_\mathrm{crit}$ we refer to
Refs.~\cite{Chiarappa:2006ae,Baron:2010bv}. 

The splitting term in the heavy doublet \eq{eq:Dsc} introduces parity and
flavour mixing between strange and charm quarks which would render the present analysis very complicated. For this reason we rely in this paper on a mixed-action
approach for the strange quark: in the valence sector we use the
so-called Osterwalder-Seiler (OS)
discretisation~\cite{Frezzotti:2004wz} with Dirac operator
\begin{equation}
  \label{eq:DOS}
  D_s^\pm = D_\mathrm{W} + m_0 \pm i \mu_s \gamma_5\,,
\end{equation}
with bare strange quark mass $\mu_s$. Formally, this introduces two
valence strange quarks with $\pm\mu_s$ as bare quark mass. We will denote these two as $s^\pm$ and they will coincide in the continuum
limit. Hence, observables computed using the one or the other will
differ by $\mathcal{O}(a^2)$ lattice artefacts. It was shown in
Ref.~\cite{Frezzotti:2004wz} that $\mathcal{O}(a)$-improvement stays
intact when $m_0$ is set to the same value $m_\mathrm{crit}$ as used
in the unitary sector. For each $\beta$-value, we choose a set of three bare
strange quark masses $a\mu_s$ as listed in \tab{tab:mus}.
The mass values are chosen such as to bracket the physical strange quark
mass independently of the light quark mass.

We remark here that in twisted mass lattice QCD the quark masses
renormalise multiplicatively with $1/Z_P$~\cite{Frezzotti:2000nk}. Since OS and
unitary actions agree in the chiral limit, also the OS strange quark
mass renormalises multiplicatively with $1/Z_P$. 
\begin{table}[t!]
 \centering
 \begin{tabular*}{.7\textwidth}{@{\extracolsep{\fill}}lrrr}
  \hline\hline
  $\beta$ & $1.90$ & $1.95$ & $2.10$ \\
  \hline\hline
  $a\mu_s$ & 0.0185 & 0.0160 & 0.013/0.0115\\
           & 0.0225 & 0.0186 & 0.015\\
           & 0.0246 & 0.0210 & 0.018\\
  \hline\hline
 \end{tabular*}
 \caption{Values of the bare strange quark mass $a\mu_s$ used for the
   three $\beta$-values. The lightest strange quark mass on the ensemble D30.48 is $a\mu_s = 0.0115$
   instead of $a\mu_s = 0.013$.}
 \label{tab:mus}
\end{table}

\subsection{Lattice Operators and Correlation Functions}

For the charged pion we use the interpolating operator
\begin{equation}
  \mathcal{O}_\pi(t) = \sum_\mathbf{x} ~\bar u(\mathbf{x},t)\, i\gamma_5\,
  d(\mathbf{x},t)
\end{equation}
projected to zero momentum. 
As interpolating operator with the quantum-numbers of the kaon we use
\begin{equation}
  \mathcal{O}_K(t) = \sum_\mathbf{x} ~\bar s^+(\mathbf{x},t)\, i\gamma_5\,
  d(\mathbf{x},t)
\end{equation}
projected to zero momentum. We use the combination of a strange quark 
with $+|\mu_s|$ and the down quark with $-|\mu_\ell|$, because it is known that observables
employing this combination are subject to milder lattice artefacts compared 
to the combination with same signs. The corresponding two-point function reads
\begin{equation}
  C_K(t-t')\ =\ \langle
  \mathcal{O}_K(t)\ \mathcal{O}^{\dagger}_K(t')\rangle
\end{equation}
and likewise the pseudo-scalar two point function $C_\pi$ with
$\mathcal{O}_K$ replaced by $\mathcal{O}_\pi$.
From the behaviour of $C_K$ ($C_\pi$) at large Euclidean time 
\begin{equation}
  \label{eq:mk}
  C_K\ \propto\ \frac{1}{2}\left( e^{-M_K t} + e^{-M_K(T-t)}\right)\,,
\end{equation}
the kaon mass $aM_K$ ($aM_\pi$) can be extracted. In order to compute
the finite volume energy shift $\delta E =E_{KK} 
-2M_K$, needed in L\"{u}schers formula to obtain the scattering length $a_0$, we have to determine the energy of the two kaon system in the
interacting case. Using the isospin $I=1$ operator 
\begin{equation}
  \mathcal{O}_{KK}(t)\ =\ \sum_\mathbf{x,x'} ~\bar
  s^+(\mathbf{x},t)\, i\gamma_5\, d(\mathbf{x},t)\ \bar
  s^+(\mathbf{x'},t)\, i\gamma_5\, d(\mathbf{x'},t)
\end{equation}
one defines the correlation function
\begin{equation}
	C_{KK}(t-t')\ =\ \langle \mathcal{O}_{KK}(t)\ \mathcal{O}^{\dagger}_{KK}(t')\rangle\,.
\end{equation}
It shows a dependence on Euclidean time similar to $C_K$ with the
addition of a time independent piece, the so-called thermal pollution
\begin{equation}
  C_{KK}\ \propto \ \frac{1}{2}\left( e^{-E_{KK} t} +
  e^{-E_{KK}(T-t)}\right)\ + \mathrm{const}\,. 
\end{equation}
To determine $\delta E$ from $C_{KK}$ we use a method which was
devised in Ref.~\cite{Feng:2009ij} for the $\pi\pi$ system with 
$I=2$. In this method, we consider the ratio
\begin{equation}
  \label{eq:ratio}
  R(t+1/2) = \frac{C_{KK}(t) - C_{KK}(t+1)}{C_K^2(t) - C_K^2(t+1)}
\end{equation}
which can be shown to have the large Euclidean time dependence
\begin{equation}
  \label{eq:ratio2}
  R(t+1/2) = A\,(\cosh(\delta E\ t') + \sinh(\delta E\ t') \coth(2 E_K
  t'))\,,
\end{equation}
with $t'= t+1/2-T/2$ and amplitude $A$.

The kaon and pion masses are affected by (exponentially suppressed) finite
size effects. The corresponding ChPT corrections
$K_{M_\pi}=M_\pi(L)/M_\pi(L=\infty)$ and
$K_{M_K}=M_K(L)/M_K(L=\infty)$ were determined from the data in
Ref.~\cite{Carrasco:2014cwa} and we reuse these values, which are 
collected in Table~\ref{tab:Mpi}. From here on we only work with finite size
corrected hadron masses:
\[
aM^\ast_H := \frac{aM_H}{K_{M_H}}\,,
\]
for $H = \pi,K$ and drop the asterisk to ease the notation.

\subsection{Stochastic LapH}

As a smearing scheme we employ the so-called stochastic
Laplacian-Heaviside (sLapH) method~\cite{Peardon:2009gh,Morningstar:2011ka}. 
In this approach the quark field under
consideration is smeared with the so-called smearing matrix
\[
  S = V_SV_S^{\dagger}.
\]
The matrices $V_S$ are matrices obtained by stacking the eigenvectors of the
lattice Laplacian, 
\begin{equation}
   \widetilde{\Delta}^{ab}(x,y;U) = \sum_{k=1}^3 \Bigl\{
   \widetilde{U}_k^{ab}(x)\delta(y,x+\hat{k})  + \widetilde{U}^{ba}_k(y)^\dagger\delta(y,x-\hat{k})
   - 2\delta(x,y)\delta^{ab}\Bigr\} \,,      
  \label{eq:laplace}
\end{equation}
columnwise. The complete set of eigenvectors spans the so called
LapH-space. The indices $a$,$b$ denote different colours, the
variables $x$,$y$ 
space-timepoints and $\widetilde{U}$ (possibly smeared) $SU(3)$-gauge link
matrices. The index $S$ on $V_S$ denotes a truncation of the eigenspectrum of
$\widetilde{\Delta}$ such that excited state contaminations of the quark field
are maximally suppressed. In addition we smear the gauge fields appearing in
Eq.~\ref{eq:laplace} with 3 iterations of 2 level HYP smearing~\cite{Hasenfratz:2001hp}, with parameters $\alpha_1 = \alpha_2 = 0.62$.
To build correlation functions we denote quark lines
connecting source and sink timeslices with
\begin{equation}
  \mathcal{Q} = \mathcal{S}\Omega^{-1}\mathcal{S} =
  V_\mathrm{s}\ (V_\mathrm{s}^\dagger \Omega^{-1} V_\mathrm{s}) 
  \ V_\mathrm{s}^\dagger \,.
  \label{eq_quarklie}
\end{equation}
where $\Omega^{-1}$ denotes the quark propagator and
$\mathcal{P}=(V_\mathrm{s}^\dagger \Omega^{-1} V_\mathrm{s})$ is called
perambulator. We use all-to-all propagators to calculate the
correlation functions which can get prohibitively expensive when done exactly.
Therefore, we employ a stochastic method with random vectors diluted
in time, Dirac-space and LapH-subspace. The all-to-all propagator then reads
\begin{equation}
   \Omega^{-1}\approx \frac{1}{N_\mathrm{R}}
 \sum_{r=1}^{N_\mathrm{R}} \sum_b X^{r[b]}\rho^{r[b]\dagger}\,,
\end{equation}
with the number of random vectors $N_\mathrm{R}$ and the compound index $r[b]$
counting the total number of random vectors and the total number of dilution
vectors $N_\textrm{D}$.
For the kaon correlation functions we reused the light quark propagators already
calculated for the $\pi\pi$ paper, Ref.~\cite{Helmes:2015gla}. The number of dilution vectors for the
light quark propagators, therefore, is the same. An exception is ensemble
D30.48 which was not included in the $\pi\pi$ paper. For this volume of $L/a=48$ the
values for the several $N_\textrm{D}$ are collected in Table~\ref{tab:dil} together with
the values of $N_\textrm{D}$ for the other lattice sizes. Concerning the newly calculated
strange quark propagators we adopted the same dilution scheme.
\begin{table}[h]
  \centering
  \begin{tabular*}{.6\textwidth}{@{\extracolsep{\fill}}crrr}
    \hline\hline
    $(L/a)^3\times T/a$     & $N_\mathrm{D}(\text{time})$ & $N_\mathrm{D}(\text{Dirac})$ &
    $N_\mathrm{D}(\text{LapH})$ \\ 
    \hline\hline
    $24^3\times48$  & 24    & 4  & 6  \\
    $32^3\times64$  & 32    & 4  & 4  \\
    $48^3\times96$  & 32    & 4  & 4  \\
    \hline\hline
  \end{tabular*}
  \caption{Summary of the number of dilution vectors, $N_\mathrm{D}$, used in
    each index. We use a block scheme in time and an interlace scheme
    in eigenvector space.}
  \label{tab:dil}
\end{table} 

An investigation of the number of random vectors $N_\mathrm{R}$ yielded no further error
reduction for the energy shift $\delta E$ when increasing $N_\mathrm{R}$ from 4
to 5 random vectors for each strange 
quark perambulator. Thus we decided to take 4 random vectors per strange quark
perambulator into account for the current analysis.

\section{Analysis Methods}
\label{sec:method}

\subsection{L{\"u}scher Method}

We are interested in the limit of small scattering momenta for the
kaon-kaon system with $I=1$ below inelastic threshold. Very much like
in the case of $\pi\pi$ scattering with $I=2$, the scattering length 
$a_0$ can be related in the finite range expansion to the energy shift
$\delta E$ by an expansion in $1/L$ as follows~\cite{Luscher:1986pf}
\begin{equation}
  \label{eq:luscher1}
  \delta E = - \frac{4 \pi a_0} {M_K L^3} \left(1 +c_1 \frac{a_0}{L}
  + c_2 \frac{a_0^2}{L^2} + c_3\frac{a_0^3}{L^3}\right) -
  \frac{8\pi^2 a_0^3}{M_K L^6}r_f +
  \mathcal{O}(L^{-7})\,,
\end{equation}
with coefficients~\cite{Luscher:1986pf,Beane:2007qr}
\[
c_1 =  -2.837297\,,\quad c_2=6.375183\,,\quad c_3=-8.311951\,. 
\]
Here, $r_f$ is the effective range parameter.
Eq.~\ref{eq:luscher1} can be solved for the scattering length $a_0/a$
given $L/a$, $a\delta E$ and $aM_K$ if the terms up to
$\mathcal{O}(1/L^5)$ are taken into account. This approach is valid only if
the residual exponentially suppressed finite volume effects are
negligible compared to the ones related for $\delta E$. Moreover,
by truncating Eq.~\ref{eq:luscher1} at $\mathcal{O}(1/L^5)$, one
assumes that the effective range has no sizeable contribution. 
We estimate the effect of this truncation in \app{sec:reff} and find it to be negligible.

\subsection{Chiral and Continuum Extrapolations}

The values of $\delta E$ and $a_0$ are calculated for each combination of
$a\mu_s$ and $a\mu_\ell$. In order to arrive at our final values for the
scattering length, we need to perform interpolations in the strange quark mass,
extrapolations in the light quark mass and the continuum
extrapolation. We adopt the following strategy: we will first
tune the renormalised strange quark 
to its physical value for all $\beta$-values and light
quark masses. Next we interpolate $M_k a_0$ in the strange quark mass
for all ensembles to this value. The value for $M_K a_0$ obtained from
this interpolation are finally extrapolated to the physical point and
the continuum limit in a combined fit. 

We use two different strategies, from here on denoted by \textbf{A} and \textbf{B}, to tune the 
renormalised strange quark mass to its physical value: 
\begin{enumerate}
\item[\textbf{A:}]  
as a strange quark mass proxy we use
\begin{equation}
  \label{eq:mesdiff}
  M_s^2 = M_K^2-M_{\pi}^2/2
\end{equation}
which is directly proportional to the strange
quark mass at leading order in ChPT. 
We interpolate $M_K a_0$ linearly in $(aM_s)^2$ to the value where
$M_s^2$ assumes its physical value for each ensemble separately. This
requires the physical value of $M_K$ and $M_\pi$ and the lattice
spacing as an input. The bare strange quark mass is not
explicitly used in this case.

\item[\textbf{B:}]  
here we are going to use the bare strange quark mass parameter $\mu_s$
explicitly. To determine the renormalised, physical value of the strange
quark mass, we first perform a global fit of the NLO SU$(2)$ ChPT
prediction for $M_K^2$
\begin{equation}
  \label{eq:mk_xpt_lat}
  (aM_K)^2 =
  \frac{P_0}{P_rP_Z} (a\mu_l+a\mu_s) \left[1 +
    P_1\frac{P_r}{P_Z}a\mu_l + \frac{P_2}{P_r^2}\right]
\end{equation}
to all our data for $aM_K$ simultaneously. Note that in SU$(2)$ ChPT
there are no chiral logarithms in $M_K^2$ predicted at NLO. 
Here we have three global fit parameters $P_0$, $P_1$ and
$P_2$.
In addition, we have $\beta$-dependent fit
parameters $P_r(\beta)$ and $P_Z(\beta)$ for $r_0/a$ and $Z_P$, respectively,
which we constrain using Gaussian priors based on the determinations of
these from Ref.~\cite{Carrasco:2014cwa}.

Hence, we have in total nine fit parameters for which we define the augmented $\chi^2$ 
function:
\begin{equation}
  \chi^2_\mathrm{aug}\ =\ \chi^2 +  \sum_\beta
  \left[\left(\frac{(r_0/a)(\beta) - P_{r}(\beta)}{\Delta
      r_0/a(\beta)}\right)^2 +
    \left(\frac{Z_{P}(\beta)-P_{Z}(\beta)}{\Delta
      Z_P(\beta)}\right)^2\right]\,. 
	\label{eq:chi_aug}
\end{equation}
Using the best fit parameters, $a\mu_s^\mathrm{ref}$ can be determined
from
\begin{equation}
  \label{eq:amu_s_phys}
  a\mu_s^\mathrm{ref} = \frac{(r_0M_K^\mathrm{phys})^2
    P_Z}{P_r P_0[1+P_1r_0m_\ell^\mathrm{phys}+P_2P_r^{-2}]} -
  \frac{P_Z}{P_r}(r_0m_\ell^\mathrm{phys}) 
\end{equation}
using the input values specified before.

This allows us to interpolate $M_K a_0$ in $a\mu_s$ to the reference value
$a\mu_s^\mathrm{ref}$ for each ensemble separately.
In the continuum limit, the physical value of the renormalised strange
quark mass, $r_0 m_s^\mathrm{phys}$, is then given by
\begin{equation}
  r_0 m_s^\mathrm{phys} =
  \frac{(r_0M_K^\mathrm{phys})^2}{P_0[1+P_1r_0m_\ell]} -
  (r_0m_\ell^\mathrm{phys})\,. 
  \label{eq:ms_phys}
\end{equation}
\end{enumerate}

\noindent In the following we will denote the combination of \textbf{M1} with
strategy \textbf{A} as \textbf{M1A} and likewise \textbf{M1B},
\textbf{M2A} and \textbf{M2B}. 

The values of $M_K a_0$ interpolated as explained above are now to be
understood at fixed renormalised strange quark mass. The quark mass
dependence of $M_K a_0$ is known from ChPT and is given at 
NLO~\cite{Gasser:1984ux,Bernard:1990kw,Chen:2006wf} by
\begin{equation}
  \label{eq:MKa0ChPT}
  M_K a_0 = \frac{M_K^2}{8\pi f_K^2}\left[-1+\frac{16}{f_K^2}\left(
    M_K^2 L^\prime -\frac{M_K^2}{2} L_5 + \zeta\right) \right]\,.
\end{equation}
Here, $L_5$ is a low energy constant (LEC) and $L^\prime$ a combination of
LECs. $\zeta$ is a known function with chiral logarithms, which can be
found in the references above. We can rewrite Eq.~\ref{eq:MKa0ChPT} in
terms of the quark masses by replacing $M_K^2$ and $f_K$ by their
corresponding LO ChPT expressions. Note that we use the convention
with $f_\pi=130\ \mathrm{MeV}$. 

As we will see later, our data for $M_K a_0$ is not sufficiently
precise to resolve terms beyond leading order, in contrast to
$M_K^2$. Including lattice 
artefacts of order $a^2$, we therefore resort to the following effective
fit ansatz for $M_K a_0$ linear in $\mu_\ell$ and $a^2$
\begin{equation}
  \label{eq:chi_cont_ext}
  M_K a_0 = Q_0 \frac{P_r}{P_Z} a\mu_\ell + Q_1 \frac{1}{P_r^2} + Q_2\,,
\end{equation}
with three free fit parameters $Q_0$, $Q_1$ and $Q_2$. 
The continuum and chiral limit for $M_K a_0$ is then given by
\[
(M_K a_0)^\mathrm{phys}\ =\ Q_0 r_0 m_\ell^\mathrm{phys} + Q_2\,.
\]
For the fit we use again an augmented $\chi^2$ like in
Eq.~\ref{eq:chi_aug} to take the errors on $r_0/a$ and $Z_P$ into
account. 

All errors are computed using the (chained) bootstrap with $1500$ bootstrap
samples. Values not determined by ourselves, e.g. for $r_0/a$ or $Z_P$
are included in the bootstrap analysis using the parametric
bootstrap. Where relevant, fits are fully correlated. The
configurations used are well separated in HMC trajectories and we have
checked explicitly for autocorrelation using a blocked bootstrap.

\section{Results}

In this section we present the results for the energy shift $\delta
E$, the scattering length $a_0$ and the chiral and continuum 
extrapolations of $M_K a_0$. From the four approaches \textbf{M1A},
\textbf{M1B}, \textbf{M2A} and \textbf{M1B} we obtain four values for
$M_K a_0$, which we combine into our final result. The spread
between the four values is used to estimate the systematic uncertainty.

\subsection{Energy Shift $\delta E$}
\label{sec:deltaE}
The energy shift is calculated from fitting Eq.\ref{eq:ratio2} to the
data of the ratio defined in Eq.\ref{eq:ratio}. Because of the
$\cosh$-like behaviour of $C_K$ and $C_{KK}$, we symmetrize the
correlation functions. For 
the kaon masses we use the results of fully correlated fits to the two-point
correlation function Eq.~\ref{eq:mk}. We repeat our fits for multiple
fit ranges for each correlation function.
The systematic uncertainties of the fitting procedures are then estimated
using the approach introduced in Ref.~\cite{Helmes:2015gla}. 
The energy value is determined as the median of the weighted distribution
over the fit ranges. The weight assigned to each fit reads 
\begin{equation}
  \label{eq:pval}
  w_X = \left[(1-2|p_X-0.5|^2)\cdot\min(\Delta X)/\Delta X\right]^2\,, 
\end{equation}
where $X=E_K,\delta E$. $p_X$ is the $p$-value of the fit and
$\Delta X$ denotes the statistical uncertainty of the considered quantity
$\braket{X}$. An estimate of the systematic uncertainty is then
calculated from the 68.54\% confidence interval of the weighted
distribution of $X$. The statistical error comes from bootstrapping
this procedure.

In order to choose the fit ranges for obtaining $M_K$ from $C_K$ and $\delta E$ from
$R$, we require several criteria to be fulfilled. Concerning the initial
timeslice $t_i$, we demand that excited states, both in $C_K$ and $R$ have
decayed away sufficiently. For $C_K$, we visually inspect the effective
mass. Since $C_K$ does not suffer from exponential error growth at late times we
set $t_f=T/2$. 
Thus we vary $t_i$ and $t_f$ within the constraints above.
In the case of the ratio, $t_f$ is set to the timeslice where $R$ starts to
deviate
significantly from
the behaviour suggested by Eq.~\ref{eq:ratio2}. The minimal number of timeslices for a fit range is chosen
with the same criterion as for $C_K$.
The values for $t_i$, $t_f$ and $t_{\textrm{min}}$ for $C_K$ and $R$ are
compiled in Tables~\ref{tab:fitrange_low}-\ref{tab:fitrange_high} for each
value of $a\mu_s$ in the Appendix~\ref{app:data}.

In Fig.~\ref{fig:dE_fit} we show exemplary fits of
the ratio Eq.~\ref{eq:ratio2} to the data for several ensembles and selected fit
ranges. At least for the Ensembles with $L=24$ the tendency of an upward bend of
the data at late times can be seen clearly.
\begin{figure}
	\centering
	\includegraphics[width=.8\textwidth]{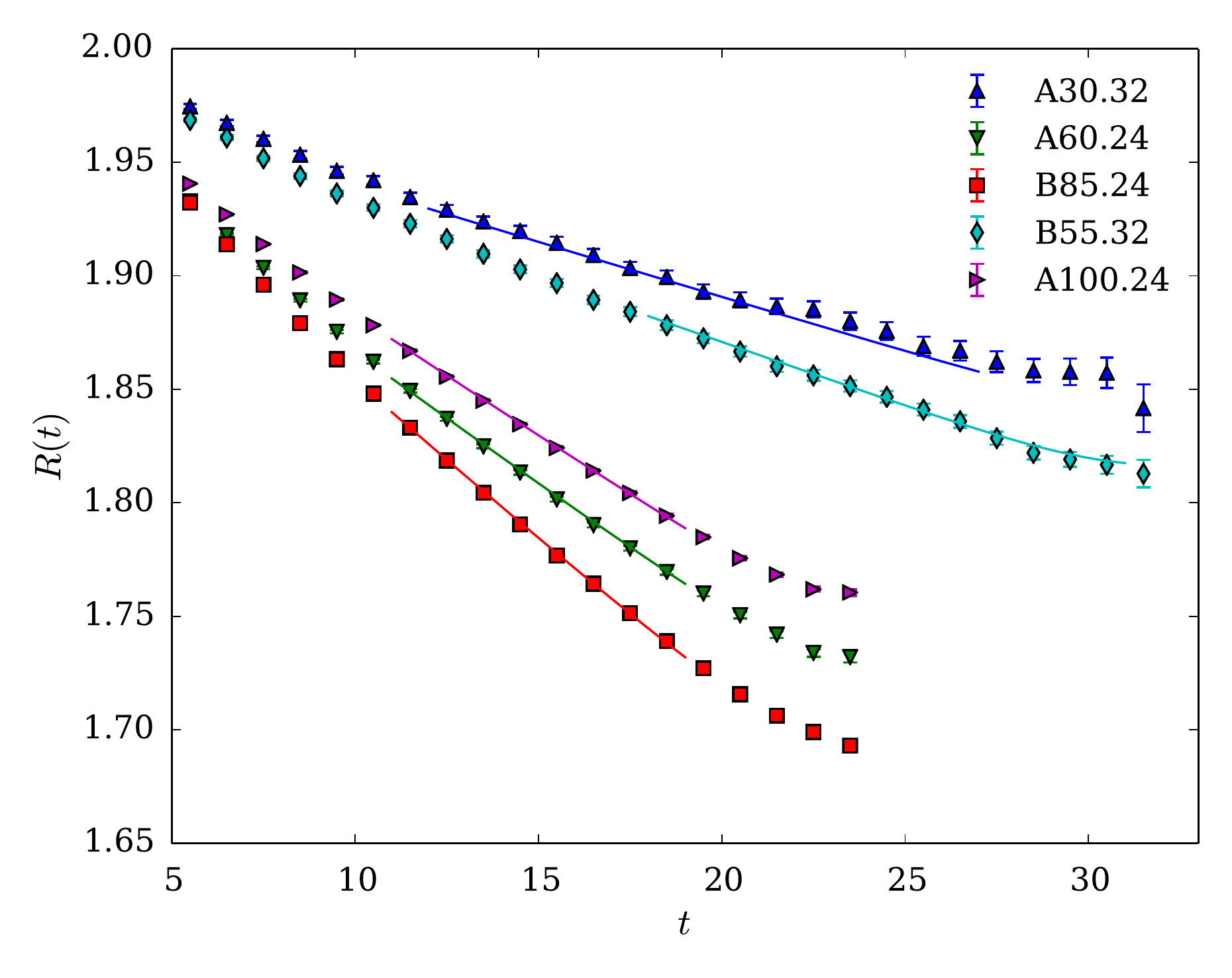}
  \caption{Representative fits of Eq.~\ref{eq:ratio2} to the ratio data for different ensembles
  at the lowest value of $a\mu_s$.}
  \label{fig:dE_fit}
\end{figure}

\begin{figure}
  \begin{subfigure}{0.48\textwidth}
    \flushleft 
    \includegraphics[width=\textwidth]{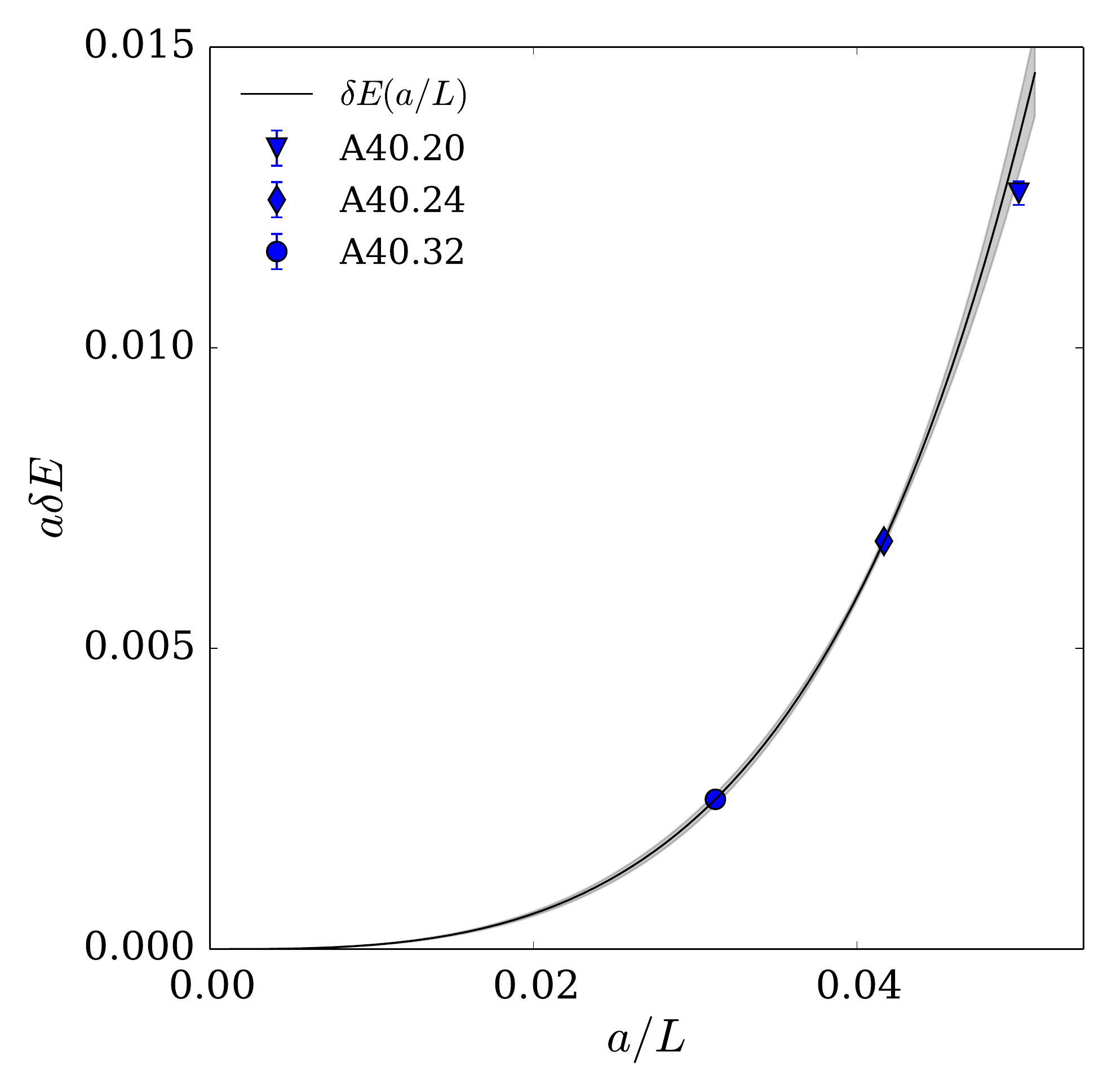}
  \end{subfigure}
  \begin{subfigure}{0.48\textwidth}
    \flushright 
    \includegraphics[width=\textwidth]{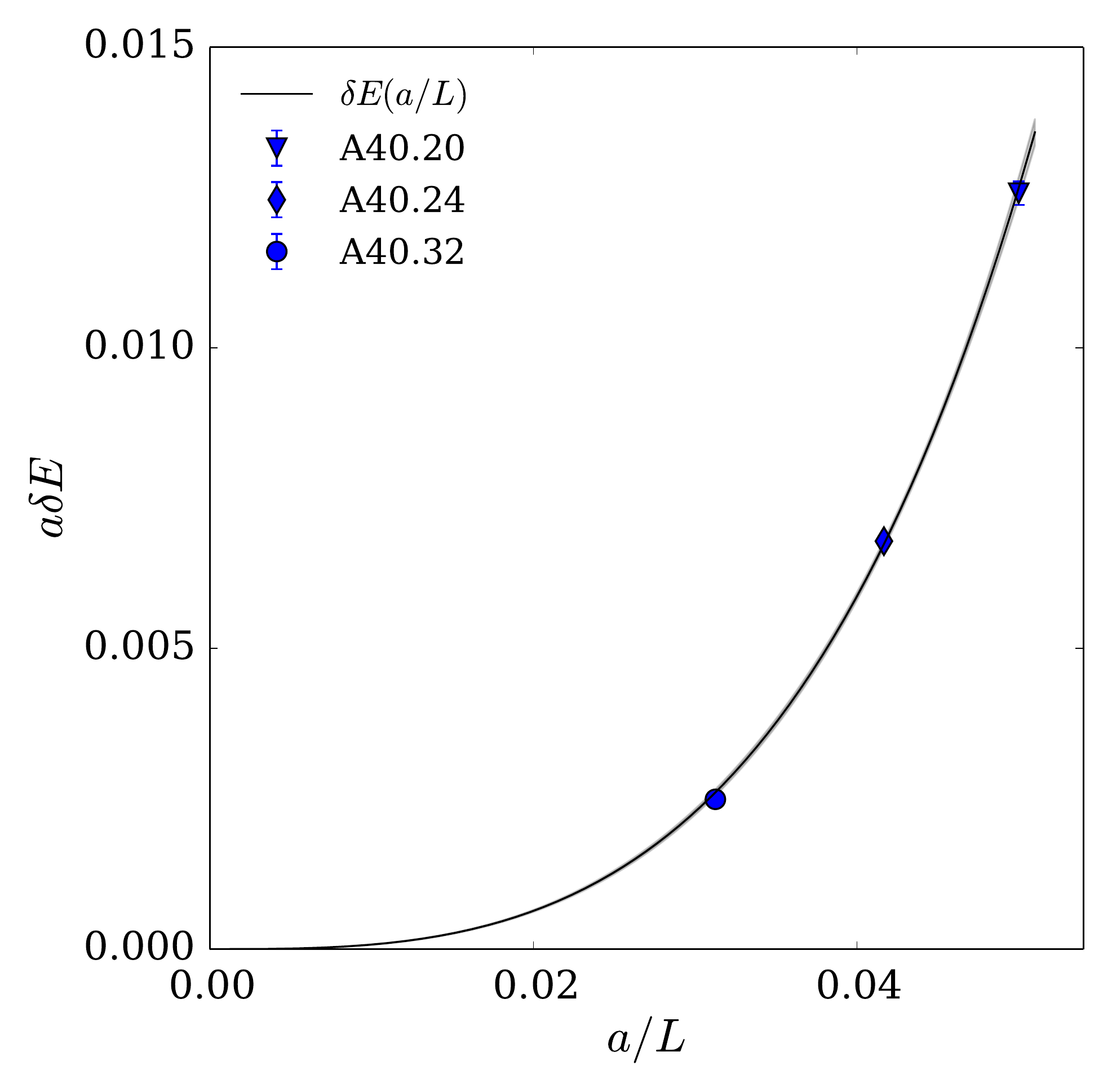}
  \end{subfigure}
  \caption{$\delta E$ as a function of $1/L$ for ensemble A40.32 with
    $a\mu_s=0.0185$. In the left panel we show as the solid line the
    solution of Eq.~\ref{eq:luscher1} for $a_0$ and $r_f$ given the
    two data points with largest $L$. In the right panel the solid
    line represents a fit of Eq.~\ref{eq:luscher1} to all three data
    points.}
  \label{fig:dE_vol}
\end{figure}

As mentioned before, for Eq.~\ref{eq:luscher1} to be valid residual
exponentially suppressed finite volume effects must be
negligible. Moreover, the terms in Eq.~\ref{eq:luscher1} of order
$1/L^6$ and higher must be negligible. 
We can test the latter for ensembles A40.20, A40.24 and A40.32, which
differ only in the volume. In Fig.~\ref{fig:dE_vol} we plot $\delta E$ 
as a function of $1/L$ for these three ensembles and
$a\mu_s=0.0185$. The other two $\mu_s$-values give similar results. 
We have solved Eq.~\ref{eq:luscher1} including all terms
up to order $1/L^6$ for $a_0$ and $r_f$ using A40.24 and A40.32 only,
the result of which is shown as the solid line with errorband in the
left panel of the figure. It leads to $M_Ka_0=-0.292(20)$. Including also
A40.20, we perform a two parameter fit with three data points finding
$M_Ka_0=-0.318(9)$. The corresponding fit is shown in
the right panel of the figure. Leaving out the effective range term at
order $1/L^6$ results in unreasonably large $\chi^2$-values.

Noting that solving Eq.~\ref{eq:luscher1} up to order $1/L^5$ for
$a_0$ for ensemble A40.32 gives $M_K a_0 = -0.315(11)$, which agrees
within error with the two estimates from above, we conclude
that $L/a=32$ is sufficiently large, while $L/a=24$ is at the
border. $L/a=20$ is certainly too small to extract $M_Ka_0$ from a
single volume neglecting the effective range term.

We checked the impact of the inclusion of $r_f$ on the extraction of $M_K a_0$
in \app{sec:reff}. With a LO ChPT estimation of $r_f$ included in the
extraction of $a_0$ the values for $M_K a_0$ vary by about one standard
deviation. The
central values for the $L/a=24$ lattices change by about 1\% on the inclusion of
the order $1/L^6$ terms (cf.~Table~\ref{tab:mka0_cmp}). Thus we attribute a conservatively estimated systematic
uncertainty of 1\% to our chiral and continuum extrapolated value of $M_K a_0$.

\subsection{Scattering Length}

Given the values of $a\delta E$ and $aM_K$, the scattering length $a_0$
is determined using Eq.~\ref{eq:luscher1}.


The number of fit ranges for extracting $a\delta E$ is low, compared to the
$\pi\pi$-case of Ref.~\cite{Helmes:2015gla}. Thus an estimate of the
systematic effects stemming from the fitting procedure is likely to be
incorrect. Therefore, instead of estimating the systematic uncertainty
introduced by the fitting procedure after the chiral
extrapolations we consider the $p$-value weighted median
over the fitranges. This procedure is further supported by the fact that
the statistical uncertainties of $M_K a_0$ do essentially not differ from the uncertainties
obtained by adding statistical and systematic uncertainties in quadrature.
The final results for $aM_K$, $a\delta E$, $a_0/a$ and $M_K a_0$ are
compiled in
Tables~\ref{tab:raw_data_mu_s_low}--\ref{tab:raw_data_mu_s_high} for
all ensembles.

\subsection{\boldmath Strategies \textbf{M1A} and \textbf{M2A}: $M_K a_0$ from fixed $M_s^2$}

To evaluate $M_K a_0$ at the physical strange quark mass, we convert 
$M_s^2$ to lattice units using $r_0/a$ listed in \tab{tab:r0values}.
First, we express $M_s^2$ in units of $r_0$ using the estimates
in \eq{eq:r0}, which gives $(r^\mathrm{M1}_0 M_s^\mathrm{phys})^2=1.33(7)$ with 
$Z_P$ from \textbf{M1} and $(r^\mathrm{M2}_0 M_s^\mathrm{phys})^2=1.34(6)$
with $Z_P$ from \textbf{M2}.
In lattice units at our three lattice spacings, these correspond to 
the values given in \tab{tab:delta_m}.

For each ensemble, we then interpolate $M_K a_0$ by performing a correlated linear
fit to the data at the three values of $a\mu_s$ (the independent variable being
$a^2 M_s^2$). An example of this is given in Fig.~\ref{fig:eval_mka0_A} in \app{app:data}.

\begin{table}
  \centering
  \begin{tabular*}{.7\textwidth}{@{\extracolsep{\fill}}lll}
    \hline\hline
    $\beta$ & $(aM_s^\mathrm{phys})^2$ (\textbf{M1}) & $(aM_s^\mathrm{phys})^2$ (\textbf{M2})\\
    \hline\hline
    1.90  &$0.0473(28)$ & $0.0475(26)$\\ 
    1.95  &$0.0400(22)$ & $0.0402(20)$\\
    2.10  &$0.0231(12)$ & $0.0232(11)$\\
    \hline\hline
  \end{tabular*}
  \caption{Physical values of $M_s^2$ for the three $\beta$
    values. The stated values correspond to the continuum values of
    $(r_0 M_s^\mathrm{phys})^2$ equal to $1.33(7)$ and $1.34(6)$ for $Z_P$ from
	\textbf{M1} and \textbf{M2}, respectively.} 
  \label{tab:delta_m}
\end{table}

\begin{figure}
  \begin{subfigure}{0.49\textwidth}
    \flushleft 
    \includegraphics[width=\textwidth]{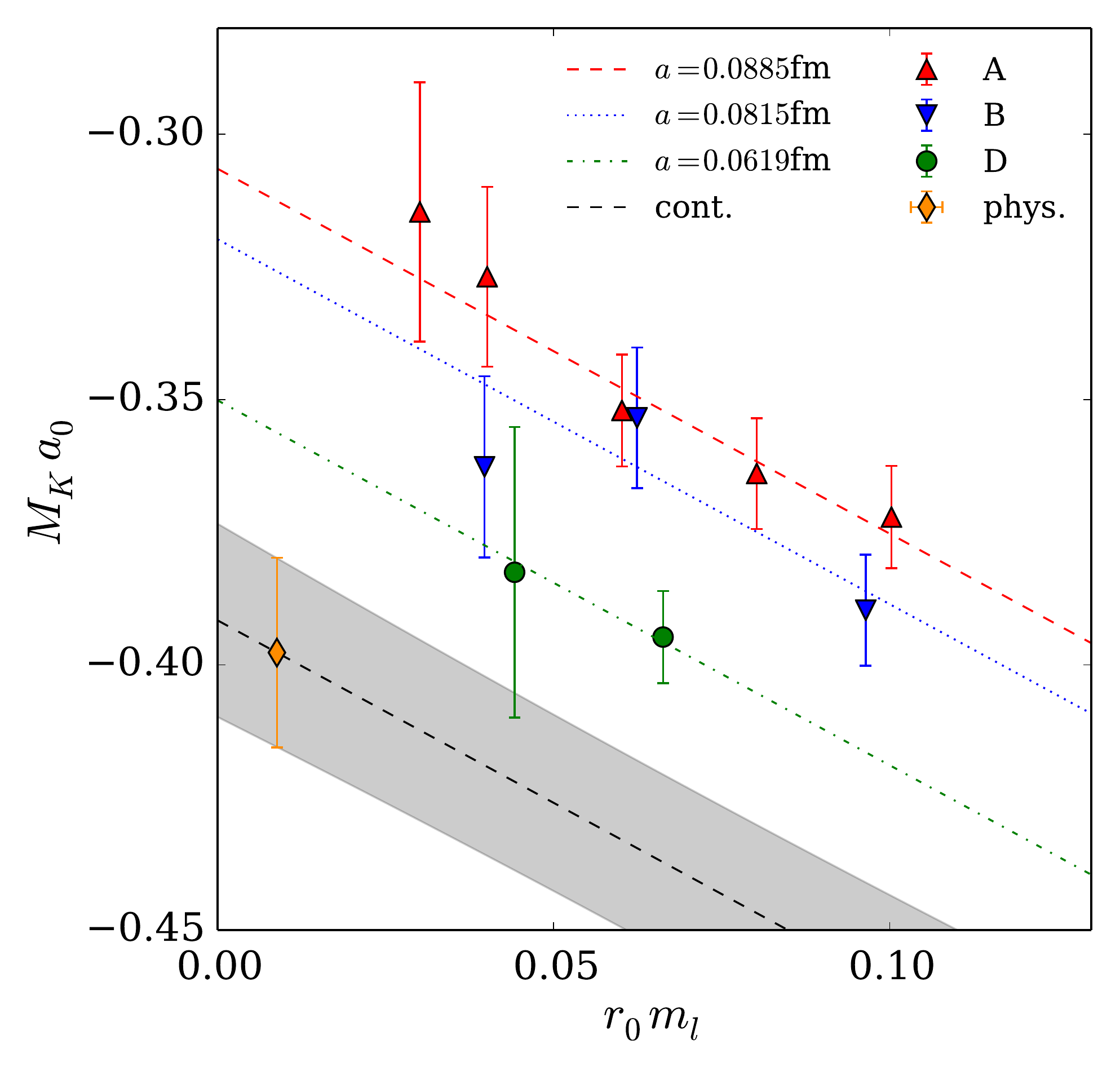}
  \end{subfigure}
  \begin{subfigure}{0.49\textwidth}
    \flushright 
    \includegraphics[width=\textwidth]{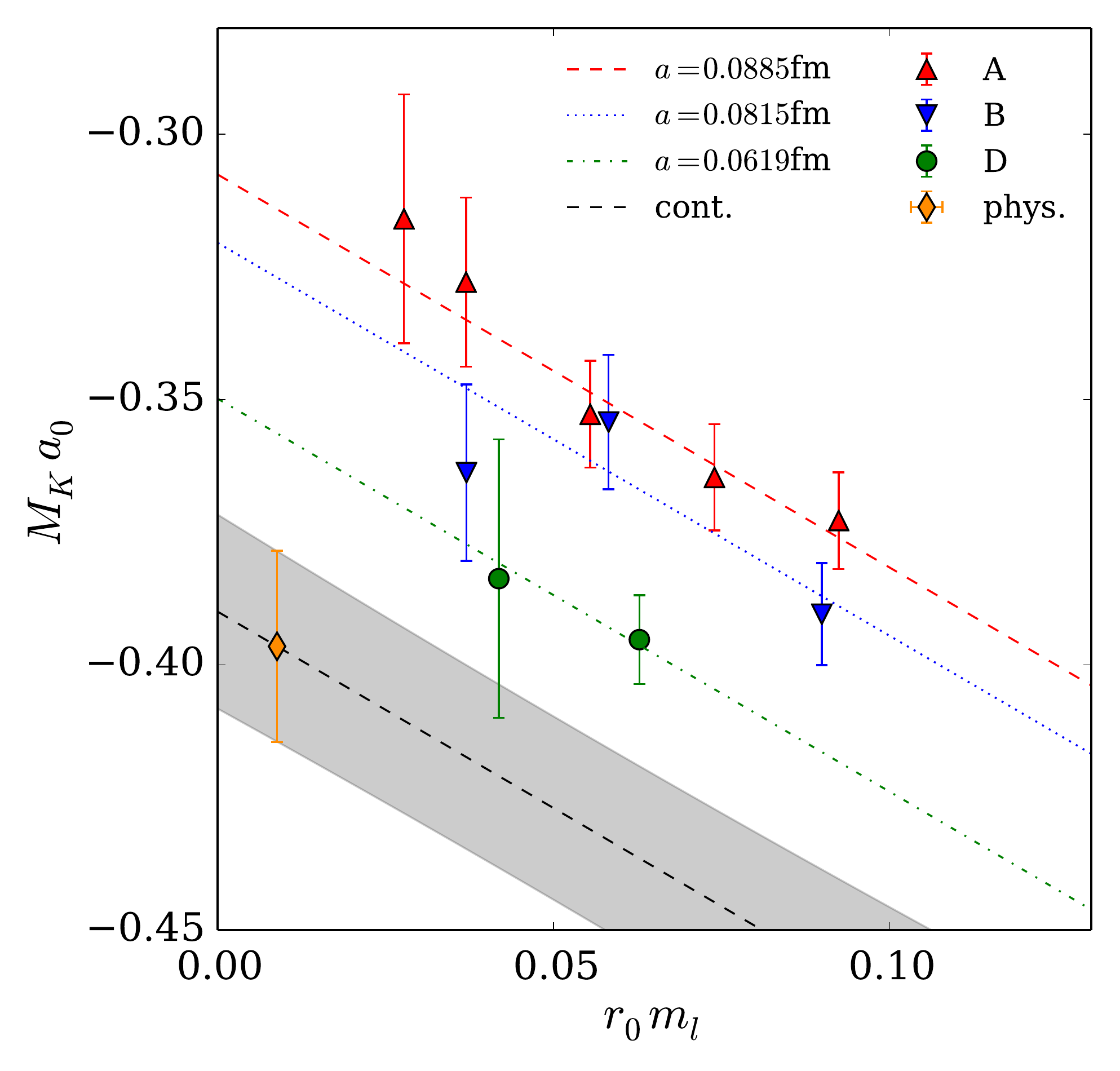}
  \end{subfigure}
  \caption{Chiral and continuum extrapolation of $M_K a_0$ to the physical
    point as a function of the light quark mass for \textbf{M1A} in
    the left and \textbf{M2A} in the right panel, respectively. Colour encoded are the three
    lattice spacings and the best fit curves. The black
    dashed line shows the continuum curve with the physical
    point result indicated by the diamond.}
  \label{fig:cont_ext_A}
\end{figure}

Having interpolated $M_K a_0$ on all ensembles, the data is extrapolated 
to the physical point and to the continuum in a global fit using
Eq.~\ref{eq:chi_cont_ext}. In Fig.~\ref{fig:cont_ext_A} the dimensionless
product $M_K a_0$ is shown as a function of $r_0
m_l$ together with the global fit for each value of $\beta$
for \textbf{M1A} in the left and for \textbf{M2B} in the right panel, 
respectively. Note that we take into account all correlation between
data which enters through the procedure for fixing the strange quark mass
at each value of the lattice spacing.
The results of the fits can be found in Table~\ref{tab:mka0_ext}.

\begin{table}
  \centering
  \begin{tabular*}{1.\textwidth}{@{\extracolsep{\fill}}lSSSS}
    \hline\hline
	& {\textbf{M1A}} & {\textbf{M2A}} & {\textbf{M1B}} & {\textbf{M2B}} \\
    \hline\hline
	$(M_Ka_0)^{\textrm{phys}}$ &-0.398(18) & -0.397(18) & -0.389(18) & -0.384(16) \\
	$\chi^2$/dof			   & 2.23{/7}    &  2.43{/7}      & 3.07{/7 }
	& 4.94{/7} \\
	$p$-value                  & 0.95       & 0.93        & 
	0.88 & 0.67 \\
    \hline
	$Q_0$ & -0.69(12)  & -0.74(12) & -0.67(12)  &
	-0.70(12) \\
	$Q_1$ & 2.4(6)    & 2.3(7)     &  2.0(6)   &
	1.7(6) \\
	$Q_2$ & -0.39(2)  & -0.39(2)   & -0.38(2)   &
	-0.38(2) \\
    \hline\hline
  \end{tabular*}
  \caption{Physical values for $M_K a_0$ obtained from the global fit
    of Eq.\ref{eq:chi_cont_ext} to the data from the different
    approaches. We also give the $\chi^2$- and $p$-values of the
    fit together with the best fit parameters
    $Q_0$-$Q_2$.}
  \label{tab:mka0_ext}
\end{table}

\subsection{\boldmath Strategies \textbf{M1B} and \textbf{M2B}: $M_K a_0$ from fixed $m_s^R$}

Analysis B involves as a first step a global fit of
Eq.~\ref{eq:mk_xpt_lat} to the values of $aM_K$.
As an example the fits to the data of the A
ensembles are shown in \app{app:data} in \fig{fig:global_fit_amu_s}
for $Z_P$ from \textbf{M1} (left panel) and \textbf{M2} (right panel),
respectively.

The fit takes into account the correlation between
data at different values of $a\mu_s$ but the same $a\mu_\ell$-value. 
The results of the global fits are compiled in Tables~\ref{tab:ms_M1}
and~\ref{tab:ms_M2}.
The fitted parameters allow us to calculate the renormalised strange quark
mass, $m_s^\mathrm{phys}$, from Eq.~\ref{eq:ms_phys}. As input, we use $r_0$ from
Eq.~\ref{eq:r0}, $Z_P$ from Table~\ref{tab:r0values}, $m_l^\mathrm{phys}$ and $M_K^\mathrm{phys}$.

\begin{table}
  \centering
  \begin{tabular*}{.7\textwidth}{@{\extracolsep{\fill}}lll}
    \hline\hline
    $\beta$ & $a\mu_s^\mathrm{ref}$ (\textbf{M1}) & $a\mu_s^\mathrm{ref}$ (\textbf{M2})\\
    \hline\hline
    1.90  & $0.0202(12)$ & $0.0204(11)$\\ 
    1.95  & $0.0182(10)$ & $0.0181(9)$\\
    2.10  & $0.0150(8)$  & $0.0151(8)$\\
    \hline\hline
    \end{tabular*}
    \caption{Values of $a\mu_s$ corresponding to the renormalized physical strange quark mass
      in lattice units for the
      three values of $\beta$ calculated from Eq.~\ref{eq:amu_s_phys}.
    }
    \label{tab:amu_s_phys}
\end{table}

For the physical values of the strange quark mass at $2\ \mathrm{GeV}$
in the $\overline{\textrm{MS}}$ scheme, we find
\begin{equation}
  \label{eq:msphys}
  \begin{split}
    m_{s}^{\textrm{phys.}} &=  101.3(4.7)\si{\mega\electronvolt}\qquad\textbf{(M1B)}\,\\
    m_{s}^{\textrm{phys.}} &=
    99.4(4.4)\si{\mega\electronvolt}\qquad\textbf{(M2B)}\,.
  \end{split}
\end{equation}
These values compare well to the corresponding results from
Ref.~\cite{Carrasco:2014cwa}: 
\begin{align*}
  m_{s}^{\textrm{ETMC}} =  101.6(4.4)\si{\mega\electronvolt}\qquad\textbf{(M1)}\,\\
  m_{s}^{\textrm{ETMC}} =  99.0(4.4)\si{\mega\electronvolt}\qquad\textbf{(M2)}\,.
\end{align*}
We can convert the values from Eq.~\ref{eq:msphys} to lattice units
for the three $\beta$-values, which we compiled in
Table~\ref{tab:amu_s_phys}. Next we interpolate $M_K a_0$ in $a\mu_s$
to these values for all ensembles. As an example we show the linear
correlated fit for ensemble B55.32 in \app{app:data} in
Fig.~\ref{fig:eval_mka0_B} for \textbf{M1B} in the left and
\textbf{M2B} in the right panel.

\begin{figure}
  \begin{subfigure}{0.49\textwidth}
    \flushleft 
    \includegraphics[width=\textwidth]{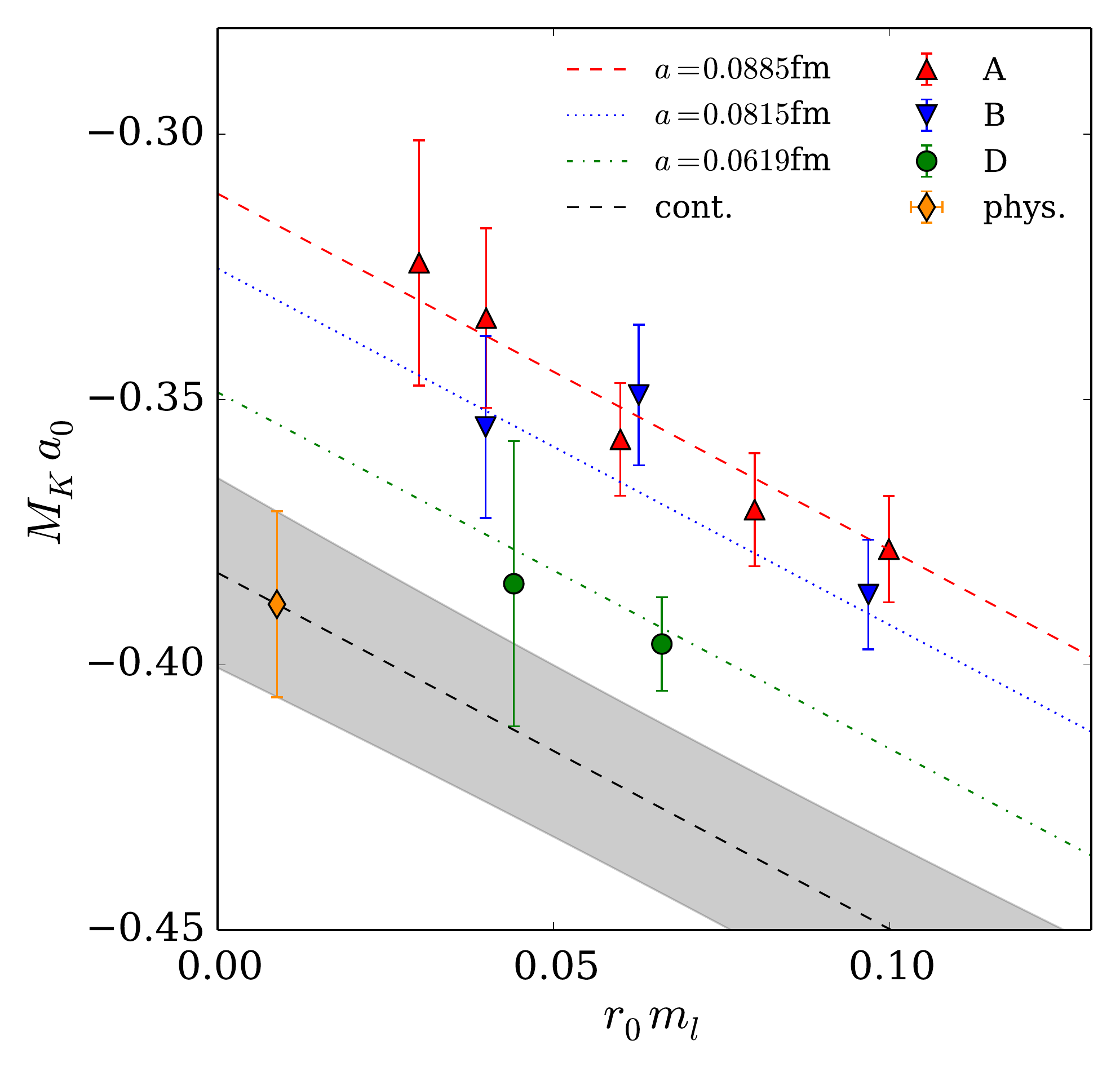}
  \end{subfigure}
  \begin{subfigure}{0.49\textwidth}
    \flushright 
    \includegraphics[width=\textwidth]{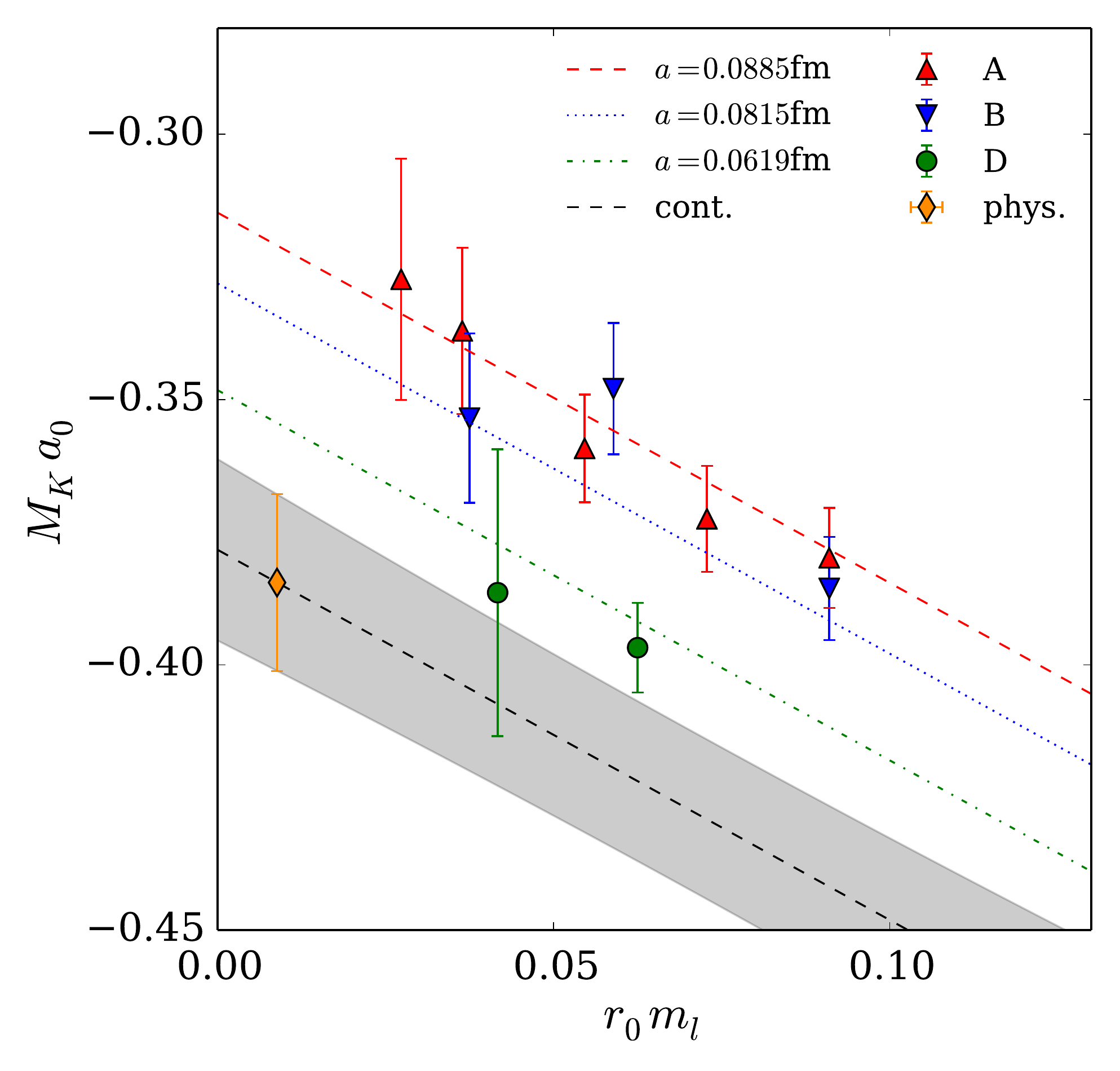}
  \end{subfigure}
  \caption{Same as Fig.~\ref{fig:cont_ext_A} but for \textbf{M1B}
    (left panel) and \textbf{M2B} (right panel).}
  \label{fig:cont_ext_B}
\end{figure}

Interpolated to the reference strange quark mass, the values of $M_K a_0$ 
are shown as a function of the renormalised light quark mass in 
Fig.~\ref{fig:cont_ext_B} in units of $r_0$. We also show the best
fit function for each $\beta$-value and the continuum extrapolation. 
The continuum extrapolated values at the physical point $(M_k
a_0)^\mathrm{phys}$ are indicated by the diamonds. Note again that due
to the strange quark mass fixing procedure all points for a single
lattice spacing are correlated.

In Table~\ref{tab:mka0_ext} we give our final results for $(M_k
a_0)^\mathrm{phys}$ for the four different approaches \textbf{M1A},
\textbf{M2A}, \textbf{M1B} and \textbf{M2B} together with the best
fit parameters $Q_{1,2,3}$, the  $\chi^2/\mathrm{dof}$- and the
$p$-value of the fit.

%

As the final result we quote the $p$-value weighted median over the
four determinations
\begin{equation}
  \label{eq:Mka0final}
  \mkaf\,.
\end{equation}
The statistical uncertainty comes from the bootstrap procedure. The
systematic uncertainty coming from the two methods to estimate $Z_P$
is estimated as follows: we 
first compute the weighted average of only \textbf{M1A} and
\textbf{M1B} and also of only \textbf{M2A} and \textbf{M2B}. The
systematic uncertainty is then taken as the deviation 
between these two weighted averages and the final result, 
Eq.~\ref{eq:Mka0final}.
For the systematic uncertainty from setting the strange quark mass we
proceed in the same way, just that we first compute the weighted
average of only \textbf{M1A} and \textbf{M2A} and also
of only \textbf{M1B} and \textbf{M2B}.
As the last error we quote the systematic uncertainty from neglecting higher order terms
in the calculation of the scattering length.
Using $M_K^{\textrm{phys}}$ we obtain for the scattering length
\begin{equation}
	\af\,.
\end{equation}

\section{Discussion}

We have used four methods to determine $M_K a_0$ at the physical light
and strange quark mass value in the continuum limit. The differences
between these methods are lattice artefacts. 
From \tab{tab:mka0_ext} it becomes clear that all four
methods give results which are well compatible within statistical
uncertainties. This gives us confidence in our procedure and in our
final result Eq.~\ref{eq:Mka0final}. The four different estimates can
still serve as an estimate of systematic effects, which are, however,
smaller than the statistical uncertainty of about 4\%. The
largest fraction of this statistical uncertainty stems from the
uncertainty in the scale.

It turns out that lattice artefacts are not negligible in $M_K a_0$:
from $\beta=1.90$ to the continuum a roughly 20\% relative change in
the result is observed. From our finest lattice spacing we still see a
change of about 8\%. It is interesting to note that our central value equals
within errors to the LO ChPT estimate
\[
(M_Ka_0)^\mathrm{LO ChPT} = -\frac{M_K^2}{8\pi f_K^2} = -0.385\,.
\]

A possibly still uncontrolled systematic uncertainty could come from
our chiral and continuum extrapolation. In lattice ChPT, usually the $a^2$
term is taken to be of higher order than the term linear in $\mu_\ell$.
For this we would need to include higher orders in the quark mass as well. However,
the precision in our data is not sufficient to resolve such terms. But
the need for the $a^2$ term is evident. Therefore, we decided to stick
to a power counting with $a^2\propto\mu_\ell$. 
An alternative and probably better chiral representation of $M_k a_0$ 
in terms of $M_K/f_K$ was used in Ref.~\cite{Beane:2007uh}. 
This representation turned out to be
not feasible for us, because we have only very little spread in
$M_K/f_K$. Smaller uncertainties on $M_K a_0$ might enable the investigation of the light and strange quark mass dependence
using mixed action ChPT at NLO.

We also cannot estimate the effects from partial quenching
of the strange quark. However, it should be noted
that the kaon masses that we obtain in the OS valence sector
at the physical strange quark mass, as set via either method \textbf{A}
or \textbf{B}, deviate from those of the unitary kaon mass published in
Ref.~\cite{OttnadPhd2014}
by a few percent at most.
Partial quenching effects in analyses
using OS valence fermions on a $N_f=2+1+1$ twisted mass sea
have been shown to be small for other observables in the past.
Moreover, we would like to remark that the dependence of
$M_Ka_0$ on $\mu_s$ is not very pronounced.
Finally, our estimate in \app{sec:reff} indicates that the $\mathcal{O}(L^{-6})$-terms in the L\"{uscher} formula Eq.~\ref{eq:luscher1} are indeed negligible for our case.
Nevertheless we do not have a sufficient number of volumes
available to determine it from the data.

\begin{figure}[t]
  \centering
  \includegraphics[width=0.8\textwidth]{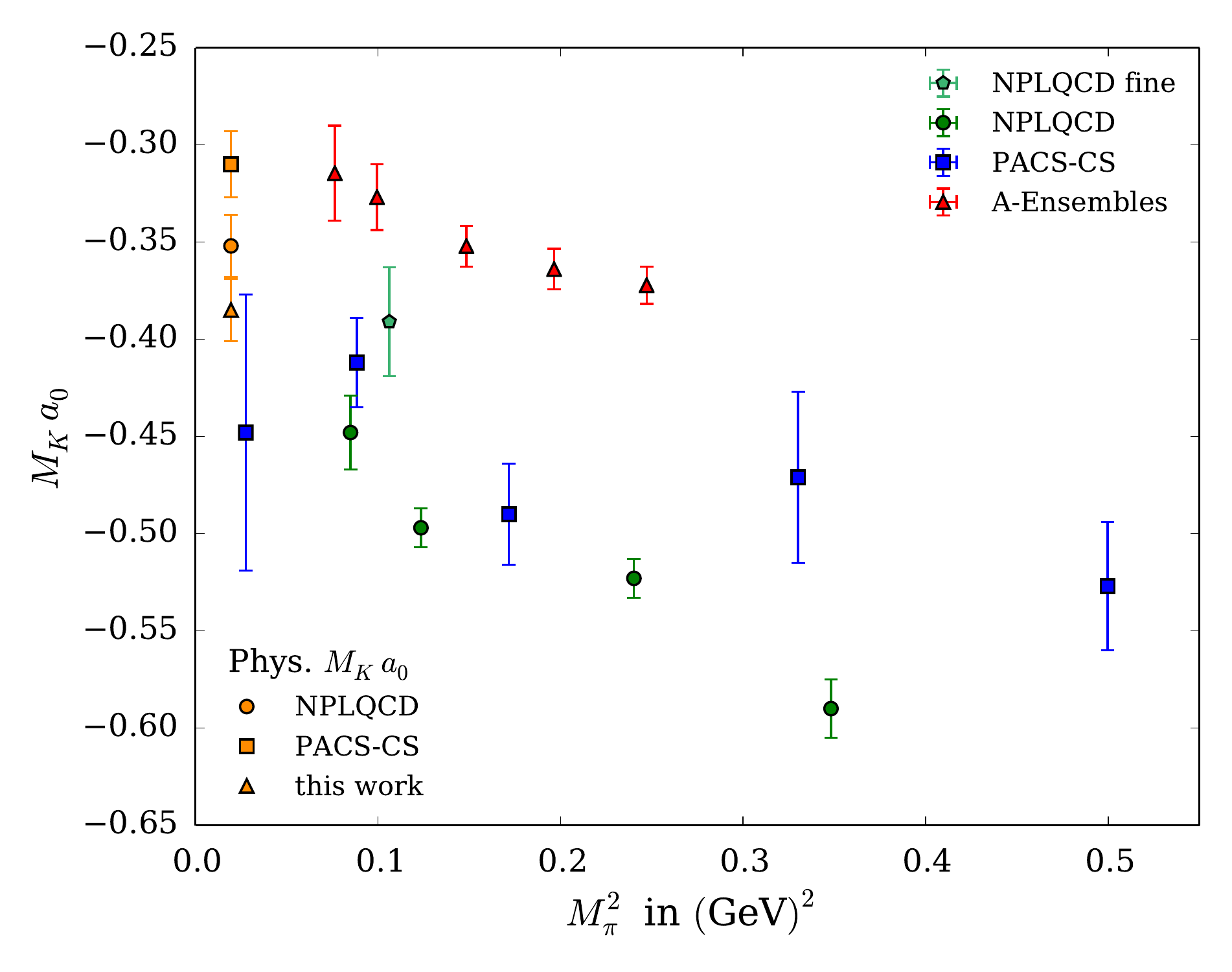}
  \caption{$M_K a_0$ as a function of $M_\pi^2$. We show our results
    at the coarsest lattice spacing value for method \textbf{M1A}
    together with the results of 
    NPLQCD~\cite{Beane:2007uh} and
    PACS-CS~\cite{Sasaki:2013vxa} with the orange circle and square indicating
    the respective final results. The orange triangle shows our final result.
  }
  \label{fig:compbare}
\end{figure}

Two other lattice calculations of $M_K a_0$ are available. The NPLQCD
collaboration used three-flavour mixed action ChPT to obtain $M_K a_0 =
-0.352(16)$, with statistical and systematic uncertainties combined in
quadrature~\cite{Beane:2007uh}. They worked with domain wall valence
quarks on a sea of $N_f=2+1$ asqtad-improved rooted staggered quarks.
A second calculation was performed
by the authors of Ref.~\cite{Sasaki:2013vxa} with $N_f=2+1$ dynamical
flavours of non-perturbatively $O(a)$-improved Wilson quarks. Their
result reads $M_K a_0 = -0.310(17)(32)$. The discrepancy between these
determinations and our final result, Eq.~\ref{eq:Mka0final}, is
quite substantial.
In the NPLQCD determination predominantly one lattice spacing of
$a=0.125\ \mathrm{fm}$ was considered in the chiral extrapolation. One
ensemble with a finer lattice spacing was included in the analysis
to attempt a quantification of discretisation errors, but it
should be noted that the uncertainty on this point was about a
factor of three larger than on all other points in the analysis.
The PACS-CS collaboration used only one lattice spacing value
with $a\sim0.09\ \mathrm{fm}$, very close to our
coarsest lattice spacing value. PACS-CS included one
ensemble with $M_\pi=170\ \mathrm{MeV}$ in their analysis, which is,
however, giving very noisy results. Both collaborations use
one strange quark mass value which was tuned to be close to physical. 

In Fig.~\ref{fig:compbare} we compare our result at the coarsest
lattice spacing, i.e. the A-ensembles, interpolated to the physical
strange quark mass with method \textbf{M1A} to the results of the other
two collaborations. There is no obvious conclusion from this
comparison. But the errors of the PACS-CS results appear to be large
enough to explain the observed differences, given the fact that the
PACS-CS result is at one lattice spacing value only. The comparison to
the NPLQCD data points is more difficult, in particular since the one
NPLQCD point with a finer lattice spacing points towards an even smaller absolute value for $M_K a_0$, though with a large statistical
uncertainty. 
This can only be resolved with continuum extrapolations
for the other formulations. However, the NPLQCD and our result agree within two
standard deviations.

\section{Summary}

We investigated the scattering length of the $K^+$-$K^+$ system
by means of finite volume methods for lattice QCD
devised by M. L\"{u}scher. The lattice formulation is Wilson twisted
mass lattice QCD at maximal twist and $N_f=2+1+1$ dynamical quark
flavours. The gauge configurations, involving 11 pion masses at 3
different lattice spacings, were generated by the ETMC. To the
author's knowledge our result represents the first study of the
$K^+$-$K^+$ system controlling lattice artefacts using three lattice
spacing values and up/down, strange and charm dynamical quarks. 
For the strange quark we used a mixed action approach with so-called
Osterwalder-Seiler valence strange quarks to be able to correct for a
slight mistuning of the sea strange quark mass value.

In total, we followed four different strategies to arrive at the
continuum extrapolated value for $M_K a_0$ at physical light and
strange quark masses. All four show very good agreement indicating
that the corresponding extrapolations are well controlled. Our final
result for the scattering length is 
\[
\mkaf
\]
from the weighted median over the four strategies. In our calculation
we find that the continuum extrapolation is vital in obtaining the
final number: from the coarsest to the continuum result we observe a
roughly 20\% difference. We think that this is also the reason for the
discrepancy we observe when comparing to the two previous lattice
calculations of $M_Ka_0$, because for the other two results a
continuum extrapolation could not be performed. 

In the near future we will extend the analysis performed here to the
pion-kaon case.

\subsection*{Acknowledgements} 

We thank the
members of ETMC for the most enjoyable collaboration. The computer
time for this project was made available to us by the John von
Neumann-Institute for Computing (NIC) on the Jureca and Juqueen
systems in J{\"u}lich. We thank A.~Rusetsky for very useful
discussions. We thank S.~Simula for the estimates of the finite size 
corrections to $M_\pi$ and $M_K$. This project was funded by the DFG
as a project in the Sino-German CRC110. The open source software
packages tmLQCD~\cite{Jansen:2009xp,Abdel-Rehim:2013wba,Deuzeman:2013xaa},
Lemon~\cite{Deuzeman:2011wz}, Eigen~\cite{eigenweb},
Boost~\cite{Gurtovoy02theboost}, SciPy~\cite{SciPy} and
R~\cite{R:2005} have been used. In addition we employed
QUDA~\cite{Clark:2009wm,Babich:2011np} for calculating
propagators on GPUs.  

\bibliographystyle{h-physrev5}
\bibliography{bibliography}

\begin{appendix}
  \section{Effective Range from ChPT}
\label{sec:reff}
We start from the partial wave expansion for the scattering amplitude $T^I(s,t,u)$~\cite{GomezNicola:2001as}
\begin{equation}
	T^I(s,t,u) = 32\pi\sum_{\ell=0}^\infty
	(2\ell+1)P_{\ell}(\cos\vartheta)t^I_{\ell}(s)\,,
\end{equation}
which depends on the Legendre polynomials $P_\ell(\cos\vartheta)$, and the
partial wave amplitudes $t^I_\ell(s)$.
The amplitudes $t^I_\ell(s)$ can be expanded in terms of the scattering momentum
$q$ and the slope parameters:
\begin{equation}
  \mathrm{Re\,}t^I_\ell = q^{2\ell}(a^I_\ell+q^2b^I_\ell+\mathcal{O}(q^4))\,.
  \label{eq:par_amp_real}
\end{equation}
Since we are interested in maximal isospin and the $s$-wave, we
take $I=1$ and $\ell=0$. This yields
\begin{equation}
  t^1_0(s)=\frac{T^1(s,t,u)}{32\pi}\,.
\end{equation}
In Ref.~\cite{GomezNicola:2001as} $T(s,t,u)$ for $K^+K^- \rightarrow K^+K^-$ is given to leading order by:
\begin{equation}
  T(s,t,u) = \frac{2M_K^2-u}{f^2_\pi}\,.
\end{equation}
To turn this into an amplitude valid for $K^+K^+$-scattering we employ crossing
symmetry which interchanges the Mandelstam variables $s$ and $u$.
With that the partial wave amplitude becomes
\begin{equation}
  t^I_0(s) = \frac{1}{32\pi}\frac{2M_K^2-s}{f^2_\pi} =
  \frac{1}{32\pi}\frac{-2M_K^2-4q^2}{f^2_\pi}\,,
  \label{eq:par_amp_exp}
\end{equation}
where we expressed $s$ with the momentum transfer $q$: $s=4(M_K^2+q^2)$.
Expanding Eq.~\ref{eq:par_amp_exp} in a Taylor series gives:
\begin{equation}
  \mathrm{Re}t^1_0(q) = -\frac{M_K^2}{16\pi f^2_\pi} - \frac{q^2}{8\pi
    f^2_\pi}\,.
  \label{eq:exp_i0s0}
\end{equation}
Comparing Eq.~\ref{eq:exp_i0s0} with Eq.~\ref{eq:par_amp_real} we can extract
$b^1_0$ and use $r_f = -2M_Kb^1_0$ to get
\begin{equation}
  r_f = \frac{M_K}{4\pi f^2_\pi}\,.
\end{equation}
To estimate the effective range we use the physical value of the kaon mass
$M_K=\SI{494.2}{\mega\electronvolt}$ and
the ChPT value $f_\pi=\SI{94.2}{\mega\electronvolt}$. Converting to a
length unit with
$\hbar c=\SI{197.37}{\mega\electronvolt\femto\metre}$ gives
\[
r_f=\SI{0.91}{\femto\metre}\,.
\]
We can use this to estimate the influence of the
$\mathcal{O}(L^{-6})$-terms on the determination of the scattering length $a_0$ from
L\"{u}scher's formula. To this end we compare the results for the scattering
length up to order $\mathcal{O}(L^{-5})$ to the ones of up to order
$\mathcal{O}(L^{-6})$ with and without the term involving the scattering length.
Table~\ref{tab:reff_scat} gives an overview over these differences .
\begin{table}[!h]
  \centering
  \begin{tabular}{lSS}
    \hline\hline
    Ensemble & {$a_0$ at $\mathcal{O}(L^{-6})$} & {$a_0$ at
      $\mathcal{O}(L^{-5})$} \\
    \hline\hline
    A60.24  &-1.405(18) & -1.393(18) \\
    A80.24  &-1.412(14) & -1.400(14) \\
    A100.24 &-1.390(12) & -1.379(12) \\
    B85.24  &-1.592(20) & -1.572(19) \\
    A40.32 & -1.350(46) & -1.346(46) \\
    D30.48  &-2.143(13) & -2.130(12) \\
    \hline\hline
  \end{tabular}
  \caption{Comparison of the scattering lengths of the $L=24$ ensembles determined
    from L\"{u}scher's formula to $\mathcal{O}(L^{-6})$, $\mathcal{O}(L^{-6})$
    without the effective range term and $\mathcal{O}(L^{-7})$, respectively at
    the lowest value of $a\mu_s$.}
  \label{tab:reff_scat}
\end{table}
For converting $r_f$ back to lattice units we use
parametric bootstrap-samples of the lattice spacing $a$. In Table~\ref{tab:mka0_cmp}
the results for $M_K a_0$ for $a_0$ up to $\mathcal{O}(L^{-6})$ and $a_0$
truncated at $\mathcal{O}(L^{-5})$ are compared.
\begin{table}[!h]
  \centering
  \begin{tabular}{lSS}
    \hline\hline
    Ensemble & {$M_K a_0$ to $\mathcal{O}(L^{-6})$} & {$M_K a_0$ to
      $\mathcal{O}(L^{-5})$} \\
    \hline\hline
    A60.24  & -0.344(5) & -0.341(4) \\
    A80.24  & -0.360(4) & -0.357(3) \\
    A100.24 & -0.367(3) & -0.364(3) \\
    B85.24  & -0.368(5) & -0.363(4) \\
    A40.32  & -0.316(11) & -0.315(11) \\
    D30.48  & -0.322(19) & -0.320(18) \\
    \hline\hline
  \end{tabular}
  \caption{Comparison of $M_K a_0$ with different orders of $L$ taken into account
    for determining $a_0$. The data shown are the $p$-value weighted medians over
    all fitranges for $\delta E$ at the lowest value of $a\mu_s$.}
  \label{tab:mka0_cmp}
\end{table}
As visible from the table the inclusion of the terms to order
$\mathcal{O}(L^{-6})$ in the determination of the scattering length does not
change the values of $M_K a_0$ beyond one standard deviation.

\section{Data Tables and Plots}
\label{app:data}

\begin{figure}
  \begin{subfigure}{0.49\textwidth}
    \flushleft 
	\includegraphics[width=\textwidth]{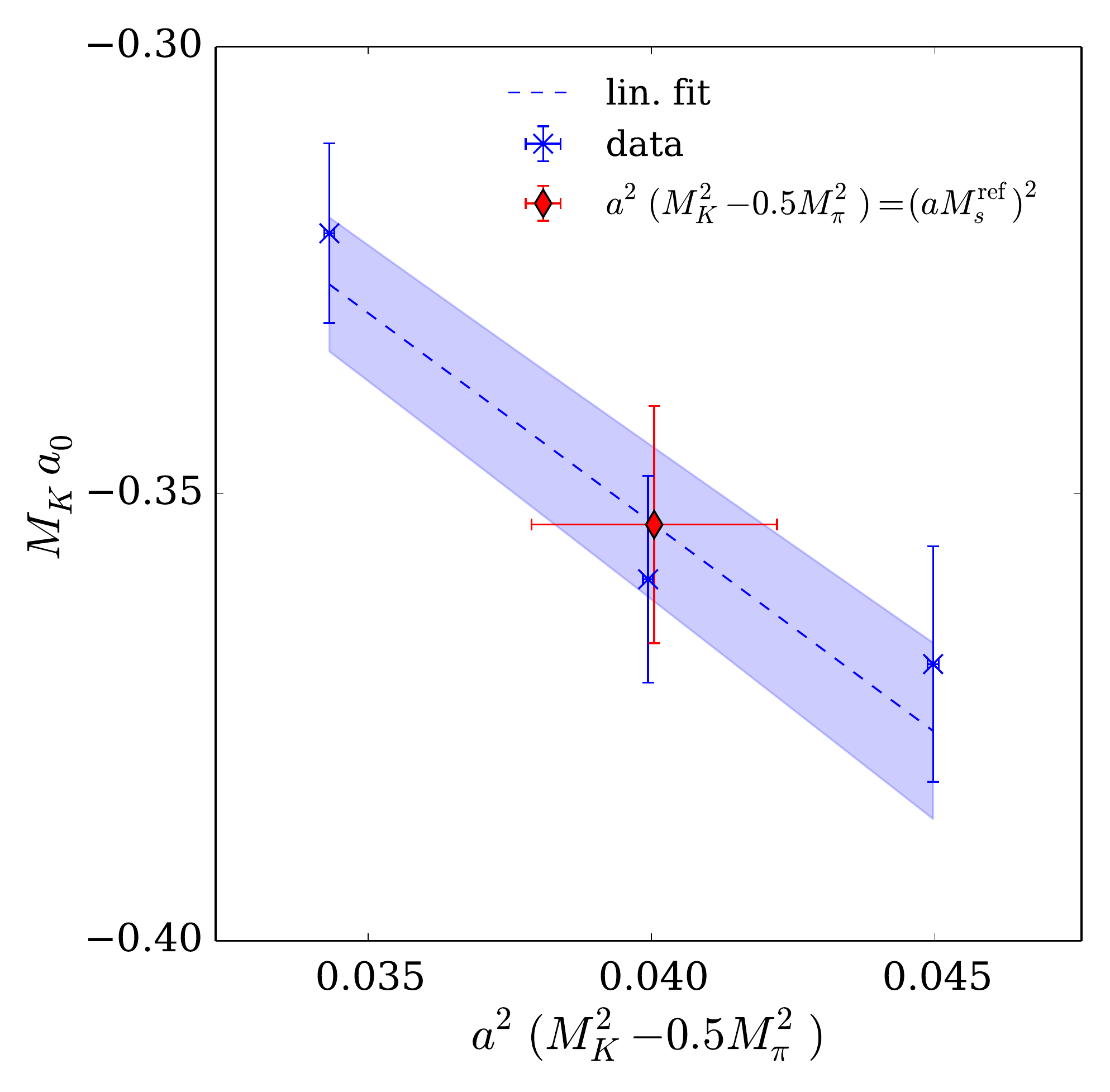}
  \end{subfigure}
  \begin{subfigure}{0.49\textwidth}
    \flushright 
	\includegraphics[width=\textwidth]{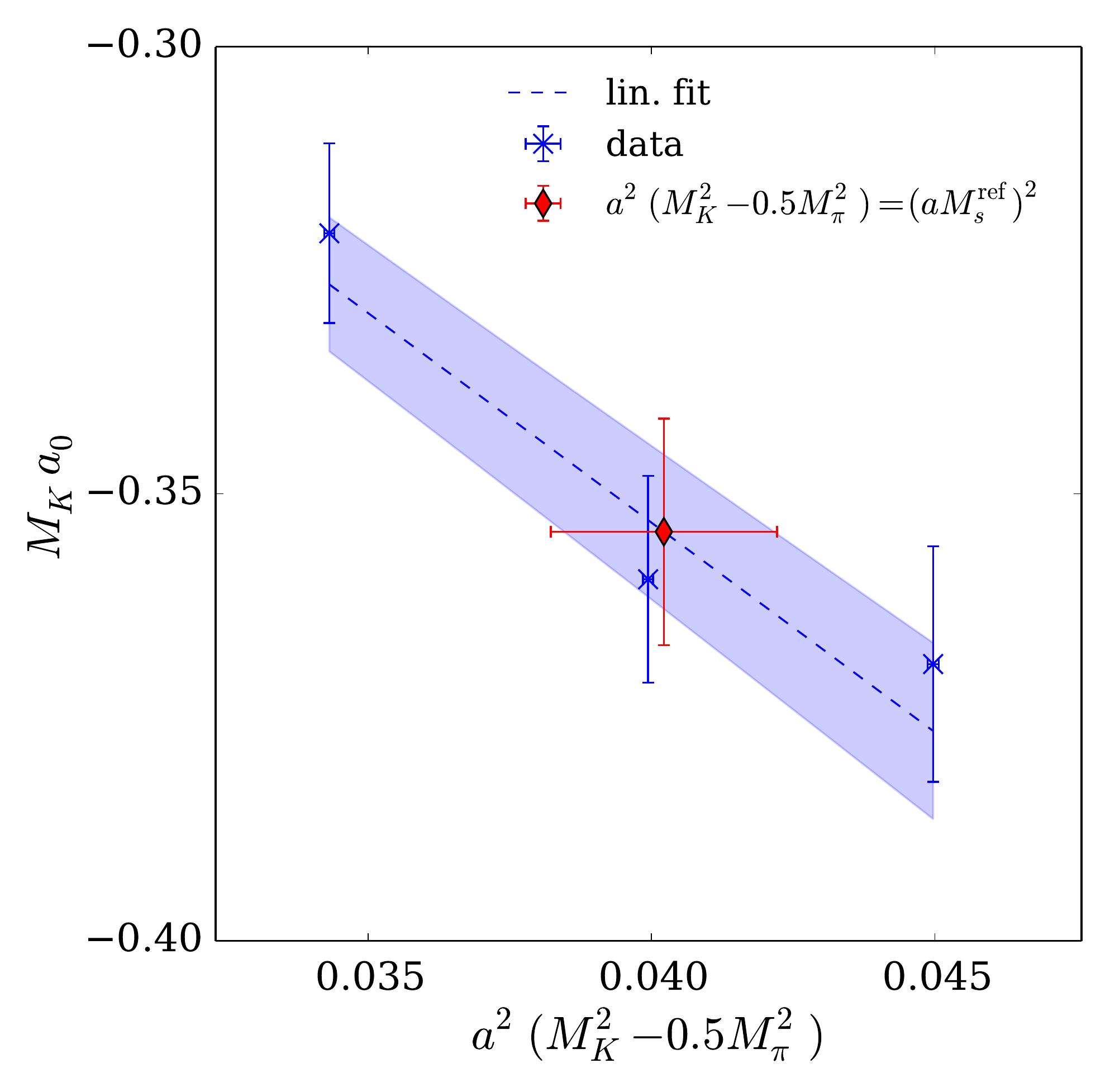}
  \end{subfigure}
  \caption{Ensemble B55.32: $M_K a_0$ as a function of $M_s^2$ for
    \textbf{M1A} in the left and \textbf{M2A} in the right panel.
    The data are
    shown as crosses. The dashed line with error band represents the
    linear fit. The interpolated value is indicated by the diamond. 
  }
  \label{fig:eval_mka0_A}
\end{figure}

\begin{figure}
  \begin{subfigure}{0.49\textwidth}
    \flushleft 
    \includegraphics[width=\textwidth]{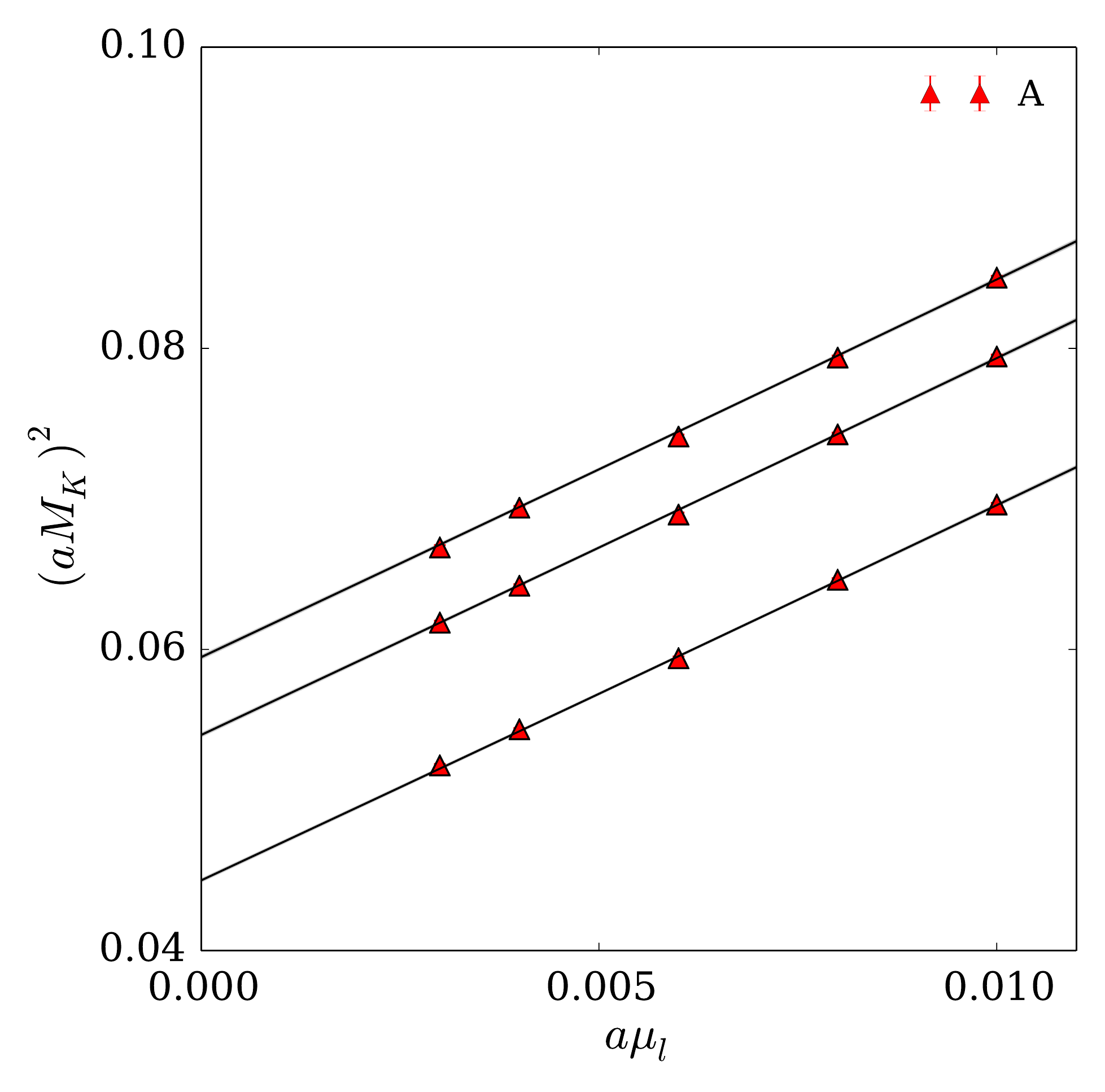}
  \end{subfigure}
  \begin{subfigure}{0.49\textwidth}
    \flushright 
    \includegraphics[width=\textwidth]{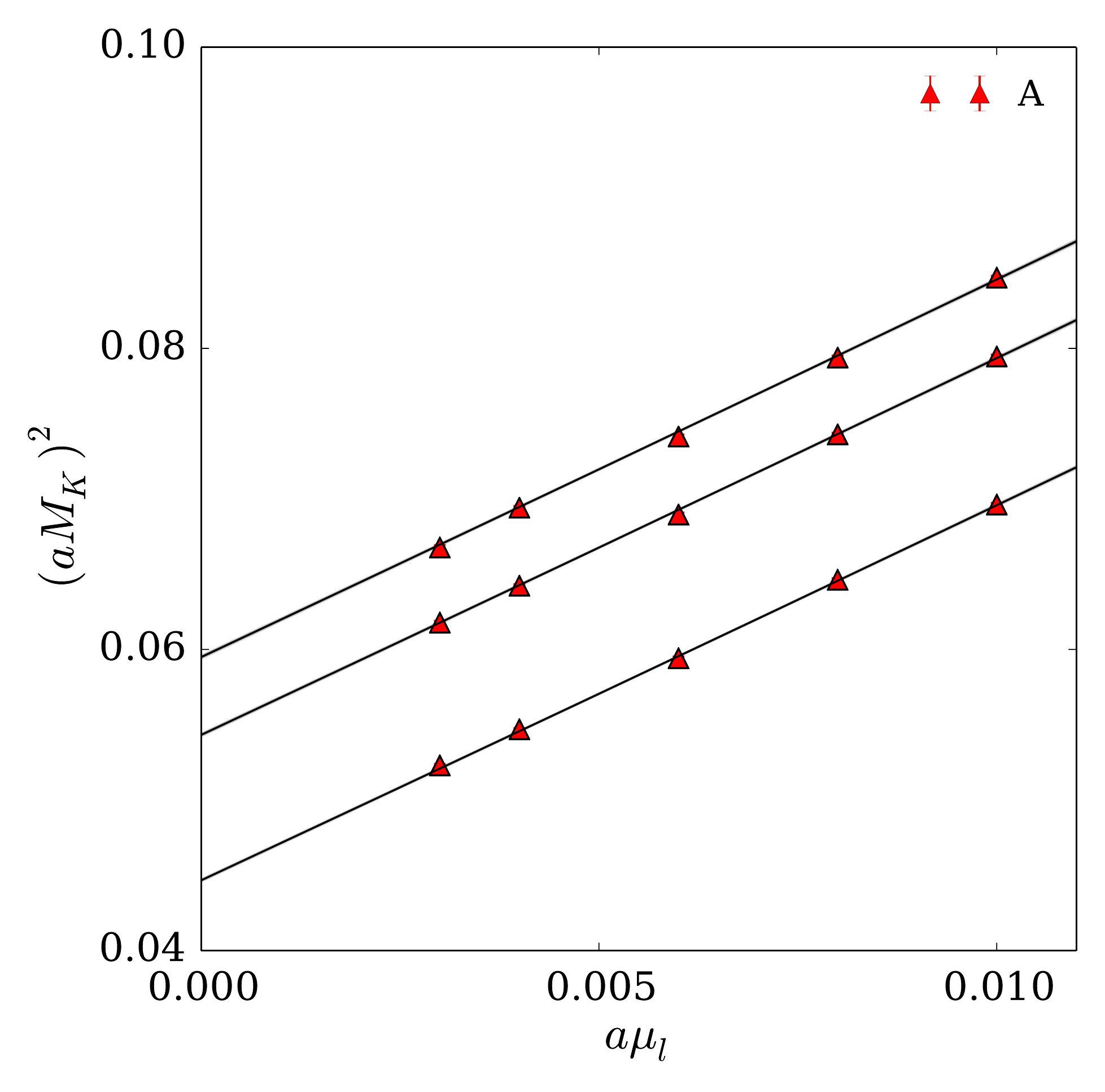}
  \end{subfigure}
  \caption{$M_K^2$ as a function of $a\mu_\ell$ for \textbf{M1B} (left
    panel) and \textbf{M2B} (right panel) for all ensembles at
    $\beta=1.90$. The lines represent the best fit of
    Eq.~\ref{eq:mk_xpt_lat} to the data. The three lines in each plot
    correspond to the three values of $a\mu_s$.}
  \label{fig:global_fit_amu_s}
\end{figure}

\begin{figure}
  \begin{subfigure}{0.49\textwidth}
    \flushleft 
    \includegraphics[width=\textwidth]{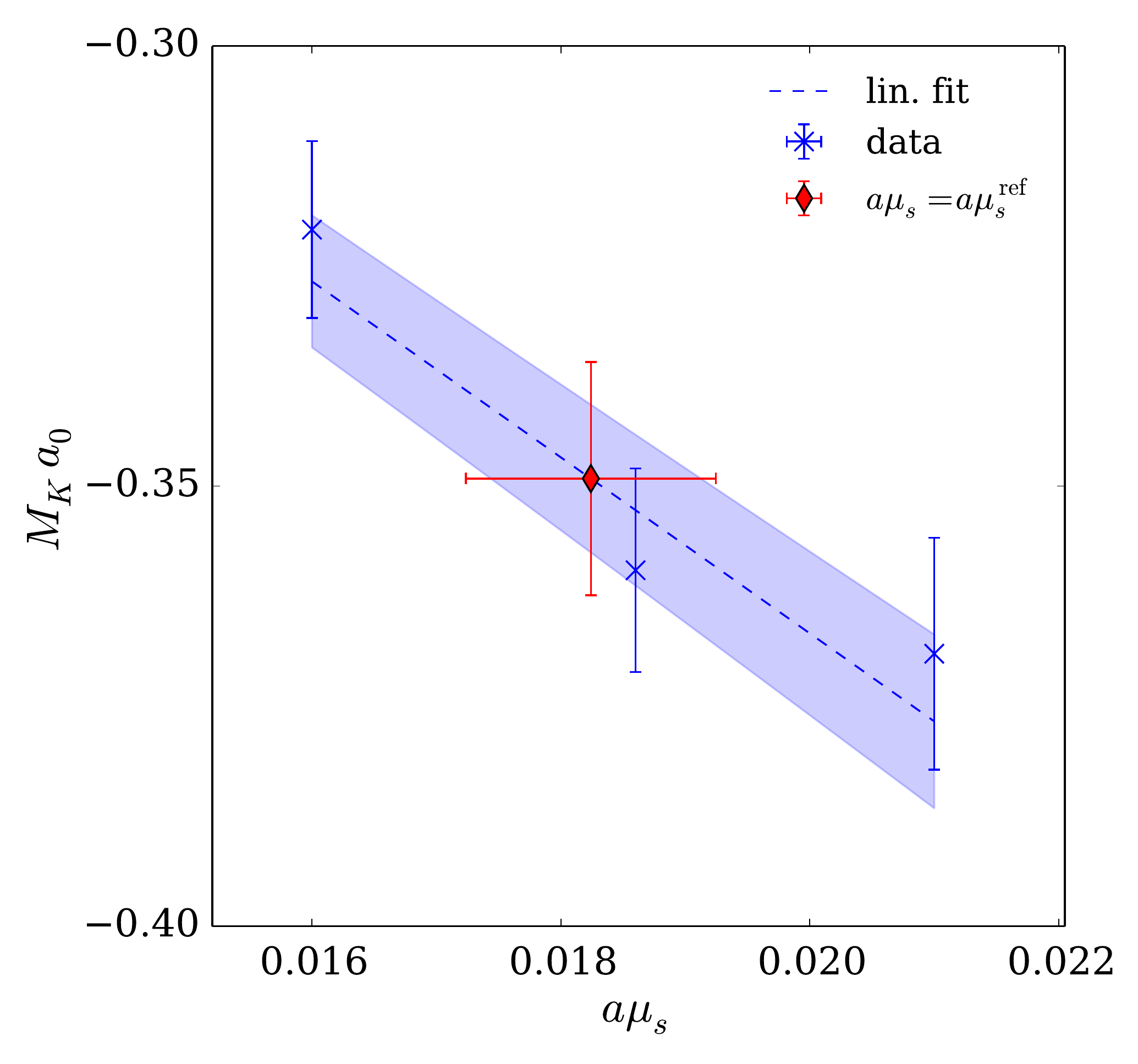}
  \end{subfigure}
  \begin{subfigure}{0.49\textwidth}
    \flushright 
    \includegraphics[width=\textwidth]{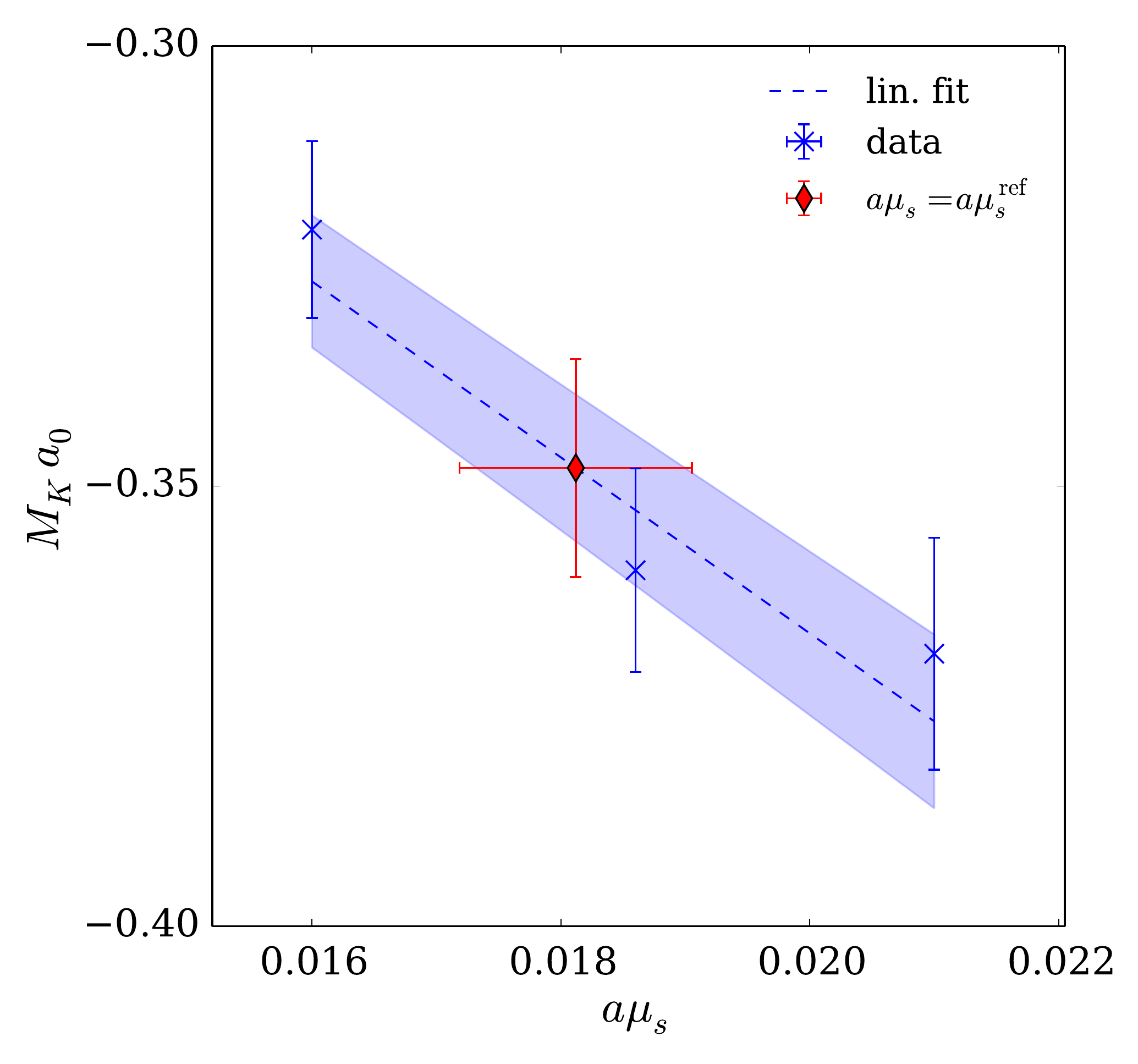}
  \end{subfigure}
  \caption{$M_K a_0$ as a function of $a\mu_s$ for ensemble B55.32 for
    \textbf{M1B} in the left and \textbf{M2B} in the right panel. 
    The line with error band represents the best fit, the interpolated
    result is indicated by the diamond.}
  \label{fig:eval_mka0_B}
\end{figure}

\begin{table}[h!]
  \centering
  \begin{tabular*}{1.\textwidth}{@{\extracolsep{\fill}}llll}
    \hline\hline
    Ensemble & $aM_\pi$ & $K_{M_\pi}$ & $K_{M_K}$ \\
    \hline\hline
    A30.32 & $0.12395(36)(14)$ & $1.0081(52)$ & $0.9954(1)$ \\
    A40.32 & $0.14142(27)(42)$ & $1.0039(28)$ & $0.9974(1)$ \\
    A60.24 & $0.17275(45)(23)$ & $1.0099(49)$ & $0.9907(1)$ \\
    A80.24 & $0.19875(41)(35)$ & $1.0057(29)$ & $0.9950(1)$ \\ 
    A100.24& $0.22293(35)(38)$ & $1.0037(19)$ & $0.9970(1)$ \\
    B35.32 & $0.12602(30)(30)$ & $1.0069(32)$ & $0.9951(1)$ \\ 
    B55.32 & $0.15518(21)(33)$ & $1.0027(14)$ & $0.9982(1)$ \\ 
    B85.24 & $0.19396(38)(54)$ & $1.0083(28)$ & $0.9937(1)$ \\
	D30.48 & $0.09780(16)(32)$ & $1.0021(7) $ & $0.9986(1)$ \\ 
    D45.32 & $0.12070(30)(10)$ & $1.0047(14)$ & $1.0000(1)$ \\
    \hline\hline                                         
  \end{tabular*}                                          
  \caption{Single pion energy levels from
	  Ref.~\cite{Baron:2010bv}, \cite{Baron:2010th} and the finite       
    size correction factors $K_{M_\pi}$ and $K_{M_K}$ computed in
    Ref.~\cite{Carrasco:2014cwa} for $M_\pi$ and $M_K$,
    respectively. The statistical uncertainty of $K_{M_K}$ is only estimated. Where not given $K_{M_K}$ is set to 1.}
  \label{tab:Mpi}
\end{table}  
\begin{table}
\centering
\begin{tabular}{llllll}
\hline\hline
&&\multicolumn{2}{c}{$C_K$}&\multicolumn{2}{c}{$R$} \\
\hline\hline
Ensemble & $a\mu_s$ &$[t_i,t_f]$ & $t_{\mathrm{min}}$ & $[t_i,t_f]$ & $t_{\mathrm{min}}$ \\ 
\hline\hline 
A30.32 & $0.0185$ & $[12,32]$ & $7$ &$ [12,32]$ & $10$ \\ 
A40.24 & $0.0185$ & $[12,24]$ & $5$ &$ [11,24]$ & $7 $ \\ 
A40.32 & $0.0185$ & $[15,32]$ & $7$ &$ [12,32]$ & $10$ \\ 
A60.24 & $0.0185$ & $[12,24]$ & $5$ &$ [11,24]$ & $7 $ \\ 
A80.24 & $0.0185$ & $[12,24]$ & $5$ &$ [11,24]$ & $7 $ \\ 
A100.24 & $0.0185$ & $[12,24]$ &$ 5$ &$ [11,24]$ &$ 7$ \\ 
\hline
B35.32 & $0.0160$ & $[15,32]$ & $7$ & $[12,32]$ & $10$ \\ 
B55.32 & $0.0160$ & $[15,32]$ & $7$ & $[14,32]$ & $10$ \\ 
B85.24 & $0.0160$ & $[12,24]$ & $5$ & $[11,24]$ & $7 $ \\ 
\hline
D30.48 & $0.0115$ & $[16,48]$ & $15$ &$[ 8,41]$ & $15$ \\ 
D45.32 & $0.0130$ & $[15,32]$ & $7$ & $[12,32]$ & $10$ \\ 
\hline\hline
\end{tabular}
\caption{Fit ranges for the lowest value of $a\mu_s$ for the kaon correlation
function $C_K$ and the Ratio $R$. The interval $[t_i,t_f]$ denotes the lowest
and largest timeslice considered in the fits, $t_{\mathrm{min}}$ is the minimal
extend of each fitrange.}
\label{tab:fitrange_low}
\end{table}
\begin{table}
\centering
\begin{tabular}{llllll}
\hline\hline
&&\multicolumn{2}{c}{$C_K$}&\multicolumn{2}{c}{$R$} \\
\hline\hline
Ensemble & $a\mu_s$ &$[t_i,t_f]$ & $t_{\mathrm{min}}$ & $[t_i,t_f]$ & $t_{\mathrm{min}}$ \\ 
\hline\hline
A30.32 & $0.0225$ &$[12,32]$ & $5$ & $[12,29]$ & $10$  \\ 
A40.24 & $0.0225$ &$[12,24]$ & $5$ & $[11,24]$ & $7 $ \\ 
A40.32 & $0.0225$ &$[15,32]$ & $7$ & $[12,32]$ & $10$  \\ 
A60.24 & $0.0225$ &$[12,24]$ & $5$ & $[11,24]$ & $7 $ \\ 
A80.24 & $0.0225$ &$[12,24]$ & $5$ & $[11,24]$ & $7 $ \\ 
A100.24 & $0.0225$ &$[12,24]$ & $5$ & $[11,24]$ & $7 $ \\ 
\hline
B35.32 & $0.0186$ &$[15,32]$ & $7$ & $[12,32]$ & $10$  \\ 
B55.32 & $0.0186$ &$[15,32]$ & $7$ & $[14,32]$ & $10$  \\ 
B85.24 & $0.0186$ &$[12,24]$ & $5$ & $[11,24]$ & $7 $ \\ 
\hline
D30.48 & $0.0150$ &$[16,48]$ & $15$ &$[ 8,41]$ & $15$  \\ 
D45.32 & $0.0150$ &$[15,32]$ & $7$ & $[12,32]$ & $10$  \\ 
\hline\hline
\end{tabular}
\caption{Same as Table~\ref{tab:fitrange_low} but for the medium value of
$a\mu_s$.}
\label{tab:fitrange_mid}
\end{table}
\begin{table}
\centering
\begin{tabular}{llllll}
\hline\hline
&&\multicolumn{2}{c}{$C_K$}&\multicolumn{2}{c}{$R$} \\
\hline\hline
Ensemble & $a\mu_s$ &$[t_i,t_f]$ & $t_{\mathrm{min}}$ & $[t_i,t_f]$ & $t_{\mathrm{min}}$ \\ 
\hline\hline
A30.32 & $0.02464$  & $[12,32]$ & $5$ & $[12,32]$ & $10$ \\ 
A40.24 & $0.02464$  & $[12,24]$ & $5$ & $[11,24]$ & $7 $ \\ 
A40.32 & $0.02464$  & $[15,32]$ & $7$ & $[12,32]$ & $10$ \\ 
A60.24 & $0.02464$  & $[12,24]$ & $5$ & $[11,24]$ & $7 $ \\ 
A80.24 & $0.02464$  & $[12,24]$ & $5$ & $[11,24]$ & $7 $ \\ 
A100.24 & $0.02464$  & $[12,24]$ & $5$ & $[11,24]$ & $7 $ \\ 
\hline
B35.32 & $0.0210 $  & $[15,32]$ & $7$ & $[12,32]$ & $10$ \\ 
B55.32 & $0.0210 $  & $[15,32]$ & $7$ & $[14,32]$ & $10$ \\ 
B85.24 & $0.0210 $  & $[12,24]$ & $5$ & $[11,24]$ & $7 $ \\
\hline 
D30.48 & $0.0180 $  & $[16,48]$ & $15$ &$[ 8,41]$ & $15$ \\ 
D45.32 & $0.0180 $  & $[15,32]$ & $7$ & $[12,32]$ & $10$ \\ 
\hline\hline
\end{tabular}
\caption{Same as Table~\ref{tab:fitrange_low} but for the highest value of
$a\mu_s$.}
\label{tab:fitrange_high}
\end{table}
\begin{table}[h!]
  \centering
  \begin{tabular*}{1.\textwidth}{@{\extracolsep{\fill}}lrrrrrr}
    \hline\hline
		$\beta$ & $P_Z$ & $P_r$ & $\bar{P}_0$ & $P_1$ & $P_2$ & $\chi^2$/dof \\ 
    \hline\hline
    $1.90$   & $0.524(7)$  & $5.22(6)$ &\multirow{3}{*}{$5.53(20)$} & \multirow{3}{*}{$0.14(3)$} & \multirow{3}{*}{$5.21(1.61)$} & \multirow{3}{*}{$6.82$}\\
    $1.95$   & $0.512(4)$  & $5.84(5)$  & & & \\
    $2.10$   & $0.516(2)$  & $7.57(8)$  & & & \\
    \hline\hline                                              
  \end{tabular*}
	\caption{Parameters from Global fit of Eq.~\ref{eq:mk_xpt_lat} to $aM_K^2$ with parameters from M1}
\label{tab:ms_M1}
\end{table}  
\begin{table}[h!]
  \centering
  \begin{tabular*}{1.\textwidth}{@{\extracolsep{\fill}}lrrrrrr}
    \hline\hline
		$\beta$ & $P_Z$ & $P_r$ & $\bar{P}_0$ & $P_1$ & $P_2$ & $\chi^2$/dof  \\
    \hline\hline
  	$1.90$   & $0.572(4)$  & $5.19(6)$ & \multirow{3}{*}{$5.65(20)$}&\multirow{3}{*}{$0.16(3)$}&\multirow{3}{*}{$7.27(1.67)$} & \multirow{3}{*}{$6.85$}\\
  	$1.95$   & $0.547(2)$  & $5.87(5)$  \\
  	$2.10$   & $0.545(2)$  & $7.56(8)$  \\
    \hline\hline                                              
  \end{tabular*}                                              
	\caption{Parameters from Global fit of Eq.~\ref{eq:mk_xpt_lat} to $aM_K^2$ with parameters from M2}
\label{tab:ms_M2}
\end{table} 
\begin{table}[h!]
  \centering
  \begin{tabular*}{1.\textwidth}{@{\extracolsep{\fill}}llllll}
\hline\hline
Ens & $a\mu_{s}$ & $aM_{K}$ & $a\delta E$ & $a_0$ & $(M_Ka_0)$ \\
\hline\hline
   A30.32 & 0.0185 & $0.2292(2)(^{+0}_{-0})$ & $0.0025(1)(^{+1}_{-0})$ & $-1.306(82)(^{+29}_{-15})$ & $-0.299(19)(^{+7}_{-3})$ \\
   A40.20 & 0.0185 & $0.2385(5)(^{+0}_{-0})$ & $0.0126(2)(^{+2}_{-1})$ & $-1.523(19)(^{+19}_{-15})$ & $-0.363(5)(^{+5}_{-3})$ \\
   A40.24 & 0.0185 & $0.2364(3)(^{+0}_{-0})$ & $0.0065(1)(^{+1}_{-0})$ & $-1.423(17)(^{+20}_{-3})$ & $-0.336(4)(^{+5}_{-1})$ \\
   A40.32 & 0.0185 & $0.2342(2)(^{+0}_{-0})$ & $0.0025(1)(^{+0}_{-0})$ & $-1.346(46)(^{+9}_{-14})$ & $-0.315(11)(^{+2}_{-3})$ \\
   A60.24 & 0.0185 & $0.2449(3)(^{+0}_{-0})$ & $0.0061(1)(^{+1}_{-0})$ & $-1.393(18)(^{+18}_{-7})$ & $-0.341(4)(^{+4}_{-2})$ \\
   A80.24 & 0.0185 & $0.2548(2)(^{+1}_{-1})$ & $0.0059(1)(^{+1}_{-0})$ & $-1.400(14)(^{+16}_{-7})$ & $-0.357(3)(^{+4}_{-2})$ \\
  A100.24 & 0.0185 & $0.2642(2)(^{+1}_{-1})$ & $0.0056(1)(^{+0}_{-0})$ & $-1.379(12)(^{+2}_{-2})$ & $-0.364(3)(^{+1}_{-1})$ \\
\hline
   B35.32 & 0.0160 & $0.2053(2)(^{+0}_{-0})$ & $0.0035(1)(^{+0}_{-0})$ & $-1.606(58)(^{+20}_{-16})$ & $-0.330(12)(^{+4}_{-3})$ \\
   B55.32 & 0.0160 & $0.2153(2)(^{+0}_{-0})$ & $0.0030(1)(^{+0}_{-0})$ & $-1.491(47)(^{+18}_{-21})$ & $-0.321(10)(^{+4}_{-4})$ \\
   B85.24 & 0.0160 & $0.2312(3)(^{+0}_{-0})$ & $0.0075(1)(^{+1}_{-1})$ & $-1.572(19)(^{+16}_{-17})$ & $-0.363(4)(^{+4}_{-4})$ \\
\hline
   D30.48 & 0.0115 & $0.1504(1)(^{+0}_{-0})$ & $0.0018(1)(^{+1}_{-1})$ & $-2.130(123)(^{+94}_{-97})$ & $-0.320(18)(^{+14}_{-15})$ \\
   D45.32 & 0.0130 & $0.1657(3)(^{+0}_{-0})$ & $0.0066(2)(^{+0}_{-2})$ & $-2.307(51)(^{+12}_{-51})$ & $-0.382(9)(^{+2}_{-9})$ \\
    \hline\hline                                              
  \end{tabular*}
  \caption{Lattice results for $aM_K$, $\delta E$, $a_0$ and $M_K a_0$ for the
  smallest value of $a\mu_s$ on all ensembles used in this study. The first
parentheses states the statistical uncertainty estimated from the bootstrap
samples of the quantity, the second one states the systematic uncertainty
estimated from the different fit ranges used for the correlation functions.}
\label{tab:raw_data_mu_s_low}
\end{table} 
\begin{table}[h!]
  \centering
  \begin{tabular*}{1.\textwidth}{@{\extracolsep{\fill}}llllll}
\hline\hline
Ens & $a\mu_{s}$ & $aM_{K}$ & $a\delta E$ & $a_0$ & $(M_Ka_0)$ \\
\hline\hline
   A30.32 & 0.0225 & $0.2491(3)(^{+2}_{-2})$ & $0.0025(2)(^{+1}_{-2})$ & $-1.436(97)(^{+45}_{-104})$ & $-0.358(24)(^{+11}_{-26})$ \\
   A40.20 & 0.0225 & $0.2577(5)(^{+0}_{-0})$ & $0.0120(2)(^{+0}_{-1})$ & $-1.565(19)(^{+4}_{-11})$ & $-0.403(5)(^{+1}_{-3})$ \\
   A40.24 & 0.0225 & $0.2560(3)(^{+0}_{-0})$ & $0.0062(1)(^{+0}_{-1})$ & $-1.454(18)(^{+9}_{-21})$ & $-0.372(5)(^{+2}_{-5})$ \\
   A40.32 & 0.0225 & $0.2538(2)(^{+0}_{-0})$ & $0.0024(2)(^{+1}_{-0})$ & $-1.391(86)(^{+46}_{-23})$ & $-0.353(22)(^{+12}_{-6})$ \\
   A60.24 & 0.0225 & $0.2638(3)(^{+1}_{-1})$ & $0.0060(1)(^{+1}_{-1})$ & $-1.459(16)(^{+14}_{-28})$ & $-0.385(4)(^{+4}_{-7})$ \\
   A80.24 & 0.0225 & $0.2732(2)(^{+1}_{-1})$ & $0.0057(1)(^{+1}_{-0})$ & $-1.435(14)(^{+13}_{-7})$ & $-0.392(4)(^{+4}_{-2})$ \\
  A100.24 & 0.0225 & $0.2823(2)(^{+0}_{-0})$ & $0.0054(1)(^{+0}_{-0})$ & $-1.412(12)(^{+8}_{-10})$ & $-0.398(3)(^{+2}_{-3})$ \\
\hline
   B35.32 & 0.0186 & $0.2185(2)(^{+1}_{-1})$ & $0.0032(1)(^{+1}_{-1})$ & $-1.589(62)(^{+37}_{-47})$ & $-0.347(14)(^{+8}_{-10})$ \\
   B55.32 & 0.0186 & $0.2281(2)(^{+0}_{-0})$ & $0.0031(1)(^{+0}_{-0})$ & $-1.578(51)(^{+12}_{-9})$ & $-0.360(12)(^{+3}_{-2})$ \\
   B85.24 & 0.0186 & $0.2429(3)(^{+1}_{-1})$ & $0.0074(1)(^{+1}_{-0})$ & $-1.619(14)(^{+13}_{-9})$ & $-0.393(3)(^{+3}_{-2})$ \\
\hline
   D30.48 & 0.0150 & $0.1674(1)(^{+0}_{-0})$ & $0.0018(1)(^{+1}_{-1})$ & $-2.318(135)(^{+126}_{-59})$ & $-0.388(23)(^{+21}_{-10})$ \\
   D45.32 & 0.0150 & $0.1748(2)(^{+3}_{-3})$ & $0.0061(2)(^{+3}_{-3})$ & $-2.250(49)(^{+80}_{-77})$ & $-0.393(9)(^{+14}_{-13})$ \\
    \hline\hline                                              
  \end{tabular*}
\caption{Same as Table~\ref{tab:raw_data_mu_s_low} but for the medium value of $a\mu_s$}
\label{tab:raw_data_mu_s_med}
\end{table} 
\begin{table}[h!]
  \centering
  \begin{tabular*}{1.\textwidth}{@{\extracolsep{\fill}}llllll}
\hline\hline
Ens & $a\mu_{s}$ & $aM_{K}$ & $a\delta E$ & $a_0$ & $(M_Ka_0)$ \\
\hline\hline
   A30.32 & 0.0246 & $0.2590(3)(^{+2}_{-2})$ & $0.0025(2)(^{+2}_{-1})$ & $-1.535(126)(^{+54}_{-26})$ & $-0.398(33)(^{+14}_{-7})$ \\
   A40.20 & 0.0246 & $0.2679(5)(^{+0}_{-0})$ & $0.0117(2)(^{+3}_{-0})$ & $-1.584(19)(^{+33}_{-5})$ & $-0.424(5)(^{+9}_{-1})$ \\
   A40.24 & 0.0246 & $0.2660(3)(^{+1}_{-1})$ & $0.0060(1)(^{+1}_{-1})$ & $-1.476(18)(^{+18}_{-23})$ & $-0.393(5)(^{+5}_{-6})$ \\
   A40.32 & 0.0246 & $0.2638(2)(^{+0}_{-0})$ & $0.0025(1)(^{+1}_{-0})$ & $-1.484(49)(^{+40}_{-20})$ & $-0.391(13)(^{+10}_{-5})$ \\
   A60.24 & 0.0246 & $0.2736(3)(^{+0}_{-0})$ & $0.0056(1)(^{+1}_{-0})$ & $-1.425(20)(^{+27}_{-3})$ & $-0.390(5)(^{+7}_{-1})$ \\
   A80.24 & 0.0246 & $0.2825(2)(^{+2}_{-2})$ & $0.0056(1)(^{+1}_{-1})$ & $-1.451(15)(^{+18}_{-18})$ & $-0.410(4)(^{+5}_{-5})$ \\
  A100.24 & 0.0246 & $0.2914(2)(^{+1}_{-1})$ & $0.0053(1)(^{+1}_{-0})$ & $-1.421(12)(^{+14}_{-8})$ & $-0.414(4)(^{+4}_{-2})$ \\
\hline
   B35.32 & 0.0210 & $0.2298(2)(^{+0}_{-0})$ & $0.0035(2)(^{+1}_{-1})$ & $-1.778(87)(^{+35}_{-27})$ & $-0.409(20)(^{+8}_{-6})$ \\
   B55.32 & 0.0210 & $0.2388(2)(^{+0}_{-0})$ & $0.0029(1)(^{+0}_{-1})$ & $-1.547(55)(^{+21}_{-25})$ & $-0.369(13)(^{+5}_{-6})$ \\
   B85.24 & 0.0210 & $0.2535(3)(^{+0}_{-0})$ & $0.0071(1)(^{+0}_{-1})$ & $-1.624(16)(^{+2}_{-12})$ & $-0.412(4)(^{+0}_{-3})$ \\
\hline
   D30.48 & 0.0180 & $0.1807(1)(^{+0}_{-0})$ & $0.0018(1)(^{+1}_{-0})$ & $-2.449(147)(^{+122}_{-43})$ & $-0.443(27)(^{+22}_{-8})$ \\
   D45.32 & 0.0180 & $0.1875(2)(^{+1}_{-1})$ & $0.0057(2)(^{+2}_{-1})$ & $-2.245(54)(^{+79}_{-36})$ & $-0.421(10)(^{+15}_{-7})$ \\
\hline\hline
  \end{tabular*}
\caption{Same as Table~\ref{tab:raw_data_mu_s_low} but for the highest value of $a\mu_s$}
\label{tab:raw_data_mu_s_high}
\end{table}

\end{appendix}

\end{document}